\documentclass[submission, Phys]{SciPost}
\usepackage{graphicx,amsmath}
\usepackage{amssymb}
\usepackage{epstopdf}
\usepackage{MnSymbol}

\makeatletter
\@ifundefined{textcolor}{}
{%
 \definecolor{BLACK}{gray}{0}
 \definecolor{WHITe}{gray}{1}
 \definecolor{ReD}{rgb}{1,0,0}
 \definecolor{GReeN}{rgb}{0,1,0}
 \definecolor{BLUe}{rgb}{0,0,1}
 \definecolor{CYAN}{cmyk}{1,0,0,0}
 \definecolor{MAGeNTA}{cmyk}{0,1,0,0}
 \definecolor{YeLLOW}{cmyk}{0,0,1,0}
}
\newcommand{\beq}{\begin{equation}}
\newcommand{\eeq}{\end{equation}}

\usepackage{dcolumn}% Align table columns on decimal point
\usepackage{bm}% bold math
\def\(({\left(}
\def\)){\right)}
\def\[[{\left[}
\def\]]{\right]}

\newcommand{\be}{\begin{equation}}
\newcommand{\ee}{\end{equation}}
\newcommand{\bea}{\begin{eqnarray}}
\newcommand{\eea}{\end{eqnarray}}

\begin{document}

\begin{center}{\Large \textbf{
Quenches in initially coupled Tomonaga-Luttinger Liquids:\\ a conformal field theory approach
%Quenches in two initially coupled conformal field theories
}}\end{center}

% TODO: write the author list here. Use initials + surname format.
% Separate subsequent authors by a comma, omit comma at the end of the list.
% Mark the corresponding author with a superscript *.
\begin{center}
Paola~Ruggiero\textsuperscript{1*},
Pasquale~Calabrese\textsuperscript{2,3},
Laura~Foini\textsuperscript{4},
Thierry~Giamarchi\textsuperscript{1},
\end{center}

\begin{center}
{\bf 1} DQMP, University of Geneva, 24 Quai Ernest-Ansermet, CH-1211 Geneva, Switzerland
\\
{\bf 2} SISSA  and  INFN,  Sezione  di  Trieste,  via  Bonomea  265,  I-34136,  Trieste,  Italy
\\
{\bf 3} International  Centre  for  Theoretical  Physics  (ICTP),  I-34151,  Trieste,  Italy
\\
{\bf 4} IPhT, CNRS, CEA, Universit\'{e} Paris Saclay, 91191 Gif-sur-Yvette, France
* paola.ruggiero@unige.ch
\end{center}

\begin{center}
\today
\end{center}

% For convenience during refereeing: line numbers
%\linenumbers

\section*{Abstract}
{\bf
% TODO: write your abstract here.
We study the quantum quench in two coupled Tomonaga-Luttinger Liquids (TLLs), from the off-critical to the critical regime, relying on the conformal field theory approach and the known solutions for single TLLs.
{We consider a squeezed form of the initial state, whose low energy limit is fixed in a way to describe a massive and a massless mode}, and we encode the non-equilibrium dynamics in a proper rescaling of the time.
In this way, we compute several correlation functions, which at leading order factorize into multipoint functions evaluated at different times for the two modes.
Depending on the observable, the contribution from the massive or from the massless mode
can be the dominant one, giving rise to exponential or power-law decay in time, respectively.
Our results find a direct application in all the quench problems where, in the scaling limit, there are two independent massless fields:
these include the Hubbard model, the Gaudin-Yang gas, and tunnel-coupled tubes in cold atoms experiments. }

\vspace{10pt}
\noindent\rule{\textwidth}{1pt}
\tableofcontents\thispagestyle{fancy}
\noindent\rule{\textwidth}{1pt}
\vspace{10pt}

\section{Introduction}
\label{sec:intro}
% TODO: write your article here.

In recent times the theoretical understanding of out-of-equilibrium homogeneous systems in 1D has become central in statistical and condensed matter physics, as counterpart to the enormous experimental advances brought by cold atoms~\cite{Bloch_RevUltracold,Langen_RevUltracoldAtoms}.
Several aspects have indeed been tackled by a variety of techniques, ranging from numerical methods (with particular reference to TEBD -- time evolved block decimation \cite{TEBD}-- and to DRMG -- density matrix renormalization group--~\cite{white1992density,schollwock2005density}, and its time dependent extension~\cite{daley2004time,white2004real}) to field theoretical techniques~\cite{giamarchi2004,cazalilla2016quantum,Calabrese_RevQuenches,bernard2016conformal}, integrability~\cite{Calabrese_IntroIntegrabilityDynamics,caux-2013,caux-16,essler2016quench,alba2021reviewGHDinhomQuenches,sotiriadis-2012,ra-20,d-14} and much more (see, e.g.,~\cite{Polkovnikov_ColloquiumNonEquilibrium,Cazalilla_RevUltracold,Gogolin_ReviewIsolatedSystems,eisert2015quantum,dalessio2016quantum,rv-21} as more comprehensive reviews).

Due to the complexity of generic out-of-equilibrium protocols, many studies focused on the simplified setup where the system is prepared in the ground state of some hamiltonian $H_0$ and then let evolve with a different one $H$: the famous \emph{quantum quench}.
While bringing important simplifications from a theory viewpoint, following the first remarkable example of Ref.~\cite{greiner2002quantum}, quenches have been realized in a variety of cases in cold atomic systems (see e.g. \cite{Zwierlein_2005,Hofferberth_2007,Cheneau_2012, kaufman-2016,brydges-2018,kww-06}).
The possibility of experimental realizations triggered a corresponding theoretical effort to set a framework in which to 
study quenches \cite{Calabrese_QuenchesCorrelationsPRL,Cazalilla_MasslessQuenchLL,Calabrese_QuenchesCorrelationsLong,Kollath_2007,Rigol_2009}.

While for free models, both on the lattice and in the continuum, quantum quench problems are most often analytically treatable as, e.g., reviewed in \cite{essler2016quench},
it has been understood that the powerful tools of integrability out-of-equilibrium \cite{caux-2013,caux-16} can lead to exact analytic results only for a limited
(but very interesting and experimentally relevant) class of initial states compatible with integrability~\cite{piroli2017what}.
Consequently, many interesting scenarios can be studied only with the help some approximate methods.

In this respect, when $H$ is at or close to a quantum critical point, a very powerful approach is brought by conformal methods.
Specifically, when the initial hamiltonian is massive, the problem can be tackled relying on an imaginary time path integral approach.
In particular, in (1+1)D, the problem is mapped to a boundary conformal field theory (BCFT) one.
This is the key result of the works by Calabrese and Cardy~\cite{Calabrese_QuenchesCorrelationsPRL,Calabrese_QuenchesCorrelationsLong}.
This description gives rise to exponential decay in time of correlations (with decay rate fixed by the initial mass).
In contrast, when also the initial hamiltonian is  massless, the correlations are expected to decay algebraically.
Such behavior is recovered for generic systems relying on the Tomonaga Luttinger Liquid (TLL) paradigm~\cite{Haldane_LuttingerLiquid,Luttinger_TLLiquid,giamarchi2004},
where initial and final hamiltonians are fully characterised by two parameters, known as sound velocity $u$ and Luttinger parameter $K$.
In this case, the power-law decays can be related in a simple way to initial and final Luttinger parameters only, as shown by Cazalilla in~\cite{Cazalilla_MasslessQuenchLL}
(see also~\cite{iucci2009quantum,iucci2010quantum,cazalilla2012thermalization,cazalilla2016quantum,cgc-18,dls-16,dpf-13,p-06,ps-11,rsm-12,dbz-12,bd-13,nbm-13,kkm-14,bd-15,ni-13,md-21} for some generalizations).

These results are somehow complementary, giving access to the dynamics after a quench starting from different classes of initial states.
Still, it is important to keep in mind their range of applicability, especially when aiming at describing quantitatively the non-equilibrium dynamics of realistic microscopical
models such as spin chains and quantum gases.
In fact, at equilibrium it is very well established that the TLL approximation is quantitatively correct in the low-energy/large-distance regime;
conversely because of the instantaneous nature of the quench after a  quantum quench the system has an energy in the middle of the many-body spectrum and by no means a low-energy description is justified.
Consequently, the extent to which Tomonaga Luttinger liquid theory can be used for the quantitative analysis of quench dynamics is a non trivial question. 
Of course, if the quench is near instantaneous but slow enough compared to high energy scales of the microscopic model (see e.g. \cite{Bernier_2014}), 
the field theory description remains valid.

%Most likely the quench has to be near instantaneous but slow enough compared to high energy scales of the microscopic model (see e.g. \cite{Bernier_2014})  for the field theory description to remain valid. 

We will assume that we can fully describe the system by an effective field theory.
In addition to the practical possibility of finding ramps with the proper speed for the field theory to remain valid, the theoretical 
study of the non-equilibrium dynamics of these conformal systems has its per se interest and provides very fundamental qualitative features that are difficult to get
by other means in such generality (e.g., the previously mentioned exponential and power-law decays of correlations).
Furthermore, many works attempted a detailed comparison between the CFT predictions and the actual non-equilibrium dynamics of lattice models, as e.g.,
\cite{uhrig2009interaction,karrasch2012luttinger,coira2013quantum,collura2015quantum,hu-13,phd-13,km-13,svp-14,dp-15,coser-2014} and, maybe surprisingly, it turned out that many features of the quench dynamics
are not only qualitatively but also quantitatively captured by the TLL approximation.
The effect of perturbations away from TLL model has also been analysed in~\cite{mitra2011mode,mitra2012thermalization,mitra2012time,mitra2013correlation}
using renormalization group (RG) methods showing that the above mentioned studies provided a useful starting point. 

Another crucial point is that the above-mentioned standard TLL approaches are only valid when the system under consideration consists of a \emph{single} field or, trivially, of several independent (pre- and post-quenches) modes.
On the other hand, in realistic (even 1D) systems this is not always the case: fermions usually carry spin, thus doubling the degrees of freedom,
like in the celebrated (Fermi-)Hubbard model~\cite{giamarchi2004,essler_Hubbard} as well as Gaudin-Yang gases~\cite{yang1967some,gaudin1967systeme};
more generally speaking, it is very common to end up in situations where the system at low-energy consists few coupled {\it species} of quasiparticles like
in the experimentally relevant example of two (or more) tubes of interacting cold bosonic gases which are tunnel-coupled~\cite{Andrews_FirstExperimentInterferenceCondensates,Shin_ExperimentSplitCondensates1,Shin_ExperimentSplitCondensates2, Shin_ExperimentSplitCondensates3,Schumm_ExperimentMatterWaveInterferometry,Albiez_ObservationTunnellingBosonicJosephsonJunction,Gati_RealizationBosonicJosephsonJunction, Levy_ExperimentalBosonicJosephsonEffects, Kuhnert_ExpEmergenceCharacteristicLength1d}.

However, in the presence of more {\it interacting species}, the quench problem becomes very challenging both numerically and analytically.
As examples,  we mention the numerical analysis performed in
Refs.~\cite{kleine2008spin,jk-15,mk-08,ekw-09,qkns-14,ymwr-14,roh-15,yr-17,sjhb-17,zvr-19,amlz-20} 
with a variety of methods.
The situation is even more complicated on the analytical side, where only few exact results are available  and mainly focused on the characterization of the
final steady state~\cite{piroli2019integrableI,piroli2019integrableII,mestyan2017exact} or work in some limits/approximations \cite{btc-17,id-17,sf-10}.
It is thus clear that any semi--quantitative, or even qualitative general picture for such problems would be not only useful, but highly desirable and this is the strategy
employed in some related works appeared in the literature~\cite{robinson2016motion,robinson2019light,mestyan2019spin,vanNieuwkerk_QuenchSineGordonSelfConsistentHarmonicApprox,vanNieuwkerk_TunnelCoupledBoseGasesLowEnergy,van2018projective,vannieuwkerk2020josephson}.
A great simplification occurs when the Hamiltonian has some symmetry and the problem with two degrees of freedom can be studied by introducing a suitable change of variable
leading to effectively decoupled modes.
In the case of tunnel coupled condensates this has allowed to study the quench of two identical tubes with a mass coupling between the two that is suddenly removed \cite{Gring_Prethermalization,Smith_PrethermalizationFromFullDistributions,Foini_CoupledLLSchmiedtmayerMassive,Foini_CoupledLLsMassiveMassless,kormos2016quantum,vanNieuwkerk_QuenchSineGordonSelfConsistentHarmonicApprox,horvath2019nonequilibrium}.

The study of the problem is, instead, much more complicated when no such obvious change of variable allows to reduce the problem to two decoupled modes,
see for instance \cite{vanNieuwkerk_TunnelCoupledBoseGasesLowEnergy}.
In this direction, few recent studies investigated the quench dynamics in the tunnelling coupling of two TLLs with different sound velocity and/or Luttinger parameter~\cite{Langen_UnequalLL,Kitagawa_DynamicsPrethermalizationQuantumNoise,ruggiero2021large}, aiming at understanding the effect of such ``imbalance'' between them.
Using a semiclassical approximation, the problem was solved via a Bogoliubov approach.
For those specific situations, this approximation was shown to give access to a very rich phenomenology \cite{ruggiero2021large}, with (i) the emergence of multiple lightcones, separating different decaying regimes;
(ii) a \emph{prethermal} regime eventually decaying into a \emph{quasi-thermal} one; (iii) non-trivial effects of a non-zero temperature in the initial state.
However, how much the above results depend on the specific quench considered there is not clear.
% general the obtained results are is not clear. 
%In this work we aim to address this question.

\subsection{Goal and main results}

The goal of the paper is to understand the quench dynamics of two initially coupled and generically different Luttinger liquids, which are let evolve independently, by relying 
on conformal methods where (differently from other approaches) general universal aspects are expected to clearly manifest.

In order to do this, we take advantage of the explicit solution of a particular (quadratic) quench problem, i.e., the one studied in Ref.~\cite{ruggiero2021large}, and in view of possible generalization, we focus on the form of the initial state in terms of the post-quench modes $b^{\dag}_{i,p}$ ($i=1,2$) (rather than looking at it as the vacuum of the pre-quench modes, as one would do in standard Bogoliubov approach).
%, rather than the basis which diagonalizes the initial Hamiltonian.
This turns out to have the following squeezed form:
%(those of the form in Eq.~\eqref{stateOp} below)
\begin{equation} \label{exactstate}
|\psi_0 \rangle = \frac{1}{\mathcal{N}}
 \prod_{p}e^{(b_{1,p}^{\dagger},b_{2,p}^{\dagger})\textbf{W}_p
\begin{pmatrix}
b_{1,-p}^{\dagger}\\
b_{2,-p}^{\dagger}
\end{pmatrix}}|0\rangle,
%\qquad\mathbb{A}(p)=
%\begin{pmatrix}A_{11} & A_{12}\\
%A_{21} & A_{22}
%\end{pmatrix}
\end{equation}
where $\mathcal{N}$ is a normalization, $|0\rangle$ the vacuum of the post-quench hamiltonian (two uncoupled TLLs), and $\textbf{W}_p$ is a two by two matrix containing all information the initial state. In particular, we realize that a crucial information is encoded in its low-energy expansion, i.e., $\textbf{W}_{p=0}$, and, specifically, in its eigenvalues: 
an eigenvalue equal to one will be associated to a massive mode, while an eigenvalue smaller
than one to a massless mode.
This is consistent with the coefficients of the squeezed initial states emerging in massive and massless quenches in a single TLLs (see Section~\ref{sec:old}).
%This is exactly the situation one encounters studying the leading order contribution in small momenta in the states derived from quenches in single -- massive or massless -- theories.

In order to get meaningful results for the dynamics, however, one must take into account the next-to-leading order term in $\textbf{W}_p$. This is also something that is corroborated by our understanding of the massive quench for single Luttinger liquid (cf. Eq.~\eqref{bstate}). For a generic theory of two different Luttinger liquids the next-to-leading order approximation of $\textbf{W}_p$ in \eqref{exactstate}, results in a matrix at linear order in $p$ which presents an off-diagonal term coupling the two diagonal modes.

In this work, focusing on the case of one massive and one massless mode, we show that if one is interested in the leading behavior -- the meaning of ``leading'' will be clear in the following -- for large time and space separation, one can safely disregard such coupling term and solve the quench of two decoupled modes. This comes at the price of dealing with multi-times correlation functions, which can be ultimately traced back to the different speed of sounds characterizing the two systems.
%arise from having different speed of sound in the two systems.

The generality of our approach lies in that the state in \eqref{exactstate}, with $\textbf{W}_p$ associated with one massive and one massless mode, can be seen as an effective description of the ground state of two LLs coupled by a generic RG relevant term, in the sense that it reproduces the leading order of equilibrium correlation functions, and is here assumed to be also a good starting point from the out-of-equilibrium problem. 
%In fact, this situation is expected to be associated to one massive mode (in the ``direction'' of the coupling term) and one massless mode (in the ``orthogonal'' one).
The main effect of going from a quadratic to a more complicated coupling (like a cosine term, associated with a truly interacting Hamiltonian) is expected to be a change in the value of some parameters in the matrix at the next to leading order: this is something well known in the context of a single theory and has been widely used for non-equilibrium settings (see Ref.~\cite{Calabrese_QuenchesCorrelationsLong} or Section~\ref{sec:old} for a brief discussion).
While the exact value of such parameters cannot be computed exactly in genuinely interacting theories, our interest here is in the functional dependence of correlation functions.

%The form of the initial state, together with a simple rescaling of time in the evolved operators (see Section~\ref{sec:ingredients}), can be complemented with CFT methods to get the asymptotics of correlations after the quench. In particular, one can, at leading order, reduce the problem, to two independent quenches (one massive and one massless) and for each use available results for quenches in a single CFT. How the breaking of the decoupling affect the results can be understood as well.

Altogether our method allows us to understand the underlying structure of correlations functions in terms of the two quasi-independent modes, and compute straightforwardly several of them. Specifically, we first compute the one- and two-point correlation functions of vertex operators of symmetric and antisymmetric sectors (see below for proper definitions), given by Eqs.~(\ref{complete1vertex}-\ref{massless1vertex_exp}) and Eqs.~(\ref{complete2vertex}-\ref{massless2vertex}) (respectively). 
Moreover, out of this approach, we can easily get other correlators, e.g., the correlations of density,  Eqs.~(\ref{D2_ij}-\ref{D2_ii}), and currents, Eqs.~(\ref{J2_ij}-\ref{J2_ii}), 
which, in the particular case of a quadratic Hamiltonian as the one studied in ~\cite{ruggiero2021large}, were not clearly accessible within a Bogoliouv approach.

%%Another important part of our analysis consists in discussing the role of corrections to such perfect decoupling, which is carefully done in Sec.~\ref{sec:gamma}. 
%%%On one hand this is a discussion of the limitation of such method, but on the other it clarify how to deal with them.
%%We show how, depending on the observable considered, such corrections can modify the power of power-laws decay (e.g., for correlations of vertex operators) or, instead, just affect the non-universal amplitude (e.g., for correlations of derivatives).

%{\color{red}Examples of application in Hubbard model somewhere?}

\subsection{Organization of the paper}
The paper is organized as follows.
In Section~\ref{sec:setting} we set the problem, including some reminders of the results obtained in Ref.~\cite{ruggiero2021large} within the semiclassical approximation.
In Section~\ref{sec:old} we give a brief overview of the two approaches to quantum quenches in conformal field theory for a single field.
The problem of two initially coupled TLLs is then studied relying on such results in Section~\ref{sec:ingredients} and Section~\ref{sec:correlations}. Specifically, Section~\ref{sec:ingredients} discusses in details 
%the features of the initial state, and the decoupling of operators after a rescaling of times, in the same basis. These are 
the main ingredients needed for the calculation of correlation functions, which is then carried out in Section~\ref{sec:correlations}.
We finally conclude in Section~\ref{conclusions}.
In order to keep the paper fluid to read, we chose to collect most of the calculations in four appendices.

\section{Setting of the problem} \label{sec:setting}

As mentioned in the introduction we are interested in studying the time evolution after a quench in which the post-quench low-energy physics is captured by
two {\it different} Tomonaga-Luttinger liquids  (in the CFT language, two free compact bosons).

%\subsection{Post-quench hamiltonian}

Without loss of generality we can write the post-quench hamiltonian as
\beq \label{Hfinal}
H = H_{u_1,K_1} [\theta_1,n_1] + H_{u_2,K_2} [\theta_2,n_2]
\eeq
with ($i=1,2$)
\beq \label{Hi}
H_{u_{i},K_i} [\theta_i, n_i]= \frac{u_{i}}{2\pi} \int {\rm d} x \left[ K_{i} (\nabla \theta_{i})^2 + \frac{1}{K_{i}} (\pi n_{i})^2  \right] \, ,
\eeq
with $[n_{i}(x),\theta_{j}(x')] = i \hbar \delta(x-x') \delta_{ij}$, and $\{ u_{i}, K_{i} \}$ the associated speeds of sound and TLL parameters.
A possible quadratic coupling between the two modes $i=1,2$ can be easily reabsorbed with a canonical transformation and hence we do not write it here.

For each $i=1,2$, the Hamiltonian $H_{i}$ in Eq.~\eqref{Hi} can be brought in a diagonal form
\beq \label{Hui}
H_{u_{i},K_i} = \sum_{p} u_{i} |p| b_{i, p}^{\dag} b_{i, p} \, .
\eeq
The fields $\theta_{i},n_{i}$ in \eqref{Hi} are related to the bosonic creation/annihilation operators $b_{i,p}^{(\dagger)}$ via:
\beq
\begin{array}{ll}\label{expansion_theta}
\displaystyle  \theta_{i}(x) = &  \displaystyle \frac{i}{\sqrt{L}}  \sum_{p \neq 0} \  e^{ - i p x - p /\Lambda}
\sqrt{\frac{\pi}{2 K_{i} |p|}} (   b^{\dag}_{i,p} -b_{i,-p})
%
%\\ \vspace{-0.2cm} \\
%
%&   \qquad + \,
\displaystyle + \frac{1}{\sqrt{L}} \theta_{i,0},
\end{array}
\eeq
\beq
\begin{array}{ll}
 \displaystyle n_{i}(x) = &  \displaystyle \, \frac{1}{\sqrt{L}}  \sum_{p\neq 0} e^{ - i p x - p/ \Lambda}
\sqrt{\frac{|p| K_{i}}{2 \pi}} (b^{\dag}_{i,p} +  b_{i,-p} )
%
%\\ \vspace{-0.2cm} \\
%
%&   \qquad + \,
\displaystyle + \frac{1}{\sqrt{L}} n_{i,0} ,
\end{array}
\eeq
%
%\begin{eqnarray}
%\theta_{\alpha} &=\\
%\phi_{\alpha}&=
%\end{eqnarray}
%
where $L$ is the system size, and we introduced an ultraviolet cutoff $\Lambda$ in order to ensure convergence of correlation functions. In the rest of the paper we focus on the thermodynamic limit (TDL), namely infinite system size.
%
%\subsection{Initial states} \label{sec:othermasses}

The initial state, of the form \eqref{exactstate}, is assumed to couple the two Hilbert spaces associated to $i=1,2$.
Otherwise, the problem factorizes in the initial variables and we go back (at leading order) to the dynamics of two decoupled single fields, which, as mentioned in the introduction, has been already largely addressed. Moreover we will assume that it describes a massive and a massless mode.
This is the case of the ground state of 
\beq \label{Hinitial}
H_{0}= H + \frac{g}{4\pi} \int dx \,  \left(\theta_1(x)-\theta_2(x)\right)^2 \ .
\eeq
considered in \cite{ruggiero2021large},
%This is exactly the case that three of us considered by Bogoliubov approach in \cite{ruggiero2021large} (see also the next subsection). 
%In this subsection we give some example of ground states of hamiltonians which are or can be effectively described by \eqref{exactstate}, which serve as initial state for the subsequent quench dynamics.
%discuss the main features of the class of initial states to which our methods applies.
or of a generic hamiltonian with a quadratic coupling in the fields $\theta_i$ (which can be recast in the hamiltonian \eqref{Hinitial}, at the price of having renormalized Luttinger parameters).
%Specifically, suppose we consider as initial Hamiltonian
%%
%\begin{equation}
%H=H_{u_{1},K_{1}}[\theta_{1},n_{1}]+H_{u_{2},K_{2}}[\theta_{2},n_{2}]- \frac{g}{4\pi}\int{\rm d}x(\alpha\theta_{1}-\beta\theta_{2})^{2},
%\end{equation}
%with $\alpha,\beta$  arbitrary real numbers.
%%
%Then, with the rescaling $\tilde{\theta}_{1}=\alpha\theta_{1},\tilde{n}_{1}=n_{1}/\alpha, \tilde{\theta}_{2}=\beta\theta_{2}, \tilde{n}_{2}=n_{2}/\beta$,
%we can rewrite the Hamiltonian as
%%
%\begin{equation}
%H=H_{u_{1},K_{1}/\alpha^{2}}[\tilde{\theta}_{1},\tilde{n}_{1}]+H_{u_{2},K_{2}/\beta^{2}}[\tilde{\theta}_{2},\tilde{n}_{2}]-\frac{g}{4\pi}\int{\rm d}x(\tilde{\theta}_{1}-\tilde{\theta}_{2})^{2},
%\end{equation}
%%
%which is of the form given in \eqref{Hinitial}.
%
More generally adding to the hamiltonian of two LLs a coupling term relevant in RG sense (e.g., a cosine) will open a gap in the spectrum, thus giving rise -- at equilibrium -- to exponentially decaying correlations (associated with an effective mass), plus possible powerlaw corrections (multiplicative and/or additive ones). Such behaviour can be effectively reproduced by a state of the form \eqref{exactstate} with a specific low energy expansion (mentioned in the introduction and explicitly given in Eq.~\eqref{stateOp} below).
%However, the value of the corresponding parameters are not easy to access, and can be taken as fitting parameters.
% (computing their specific value for a specific hamiltonian is not our goal here).
Crucially, 
%such a state can always be though as the ground state of two TTLs with a quadratic coupling with mass given by the effective mass.
the correlations functions of such a state at large distance are characterised only by a mass which can always be though as an effective parameter in the ground state of two TTLs.
% and in the paper we will think of the state in this term, and use in this sense the parameters introduced in Section~\ref{sub:bog}.

What is non-trivial, then, is that this ansatz for the state still correctly describes the dynamics after a quench.
In general this is not guaranteed already for a single theory, as subleading corrections might become dominant at late times. 
In the case of an interacting post-quench hamiltonian, indeed, whether a squeezed state approximation is justified in some regimes is subject of current research~\cite{sotiriadis2014validity,horvath2016initial,kst-18,kukuljan2020out,gaussification}. 
If the final hamiltonian is critical, instead, arguments of RG theory of boundary critical behaviour~\cite{diehl-86} have been used to justify a squeezed state form~\cite{Calabrese_QuenchesCorrelationsLong} (see Section~\ref{sec:CC} below). 
Since we focus on a critical final hamiltonian, we assume that this is still the case for two theories. Whether at some later time, our assumption on the initial state breaks down is an open problem, that, nevertheless, is beyond the aim of this paper.

%\subsection{Observables}

In the following sections we discuss the dynamics of the symmetric and antisymmetric modes
\begin{equation}
\theta_{\pm} =\frac{ \theta_1\pm \theta_2}{ \sqrt{2}}, 
\end{equation}
with speed of sound $u$ and TLL parameter $K$ equal and given by
\beq \label{uK_def}
u K = \frac{1}{2} (u_1 K_1+u_2 K_2), \hspace{2cm} \frac{u}{K} = \frac{1}{2} \left(\frac{u_1}{ K_1}+ \frac{u_2}{K_2}\right) \, .
\eeq
$\theta_{\pm}$ are often relevant
in systems with two types of degrees of freedom, as the Hubbard model or
the tunnel-coupled condensates.
While these variables are obviously the most appropriate ones in the case of two identical systems with coupling in $(\theta_1 - \theta_2)$,
as the initial and final Hamiltonian are decoupled in this basis and the quench occurs only
in the antisymmetric sector \cite{kardar1986josephson,gritsev2007linear}, this is not true
in general when the two systems are different.

\subsection{Reminders of Bogoliubov approach and some notations} \label{sub:bog}

%Starting from the hamiltonian~\eqref{Hinitial}, it is clear that its ground state is factorized as the product of the vacua of the pre-quench modes (the $\eta$ operators of Eq.~\eqref{HinitialD},
%i.e., $|\psi_0 \rangle = |0_{\eta_m} \rangle \otimes |0_{\eta_0} \rangle $, where the states $|0_{\eta_{m/0}} \rangle $ satisfy $\eta_{m/0,p} |0_{\eta_{m/0}} \rangle = 0, \, \forall p$).
%Then, because of the quadratic nature of both pre- and post-quench hamiltonians,  the initial state in the basis of the post-quench hamiltonian takes is squeezed state of the form \eqref{exactstate}, as anticipated.

To make contact with Ref.~\cite{ruggiero2021large}, we recall that in the Bogoliubov approach to the quench dynamics \cite{essler2016quench}, the initial state is assumed to be the ground state (or even another eigenstate \cite{bkc-14}, but we do not consider this case here)
of a quadratic hamiltonian, as, e.g., \eqref{Hinitial}.
Hence, the standard way to solve the quench dynamics of interest is to exploit its quadratic nature and
perform a Bogoliubov transformation to bring it in the following diagonal form
\beq \label{HinitialD}
H_{0} = \sum_p \lambda_{m,p} \eta_{m,p}^{\dag} \eta_{m,p} + \sum_p \lambda_{0,p} \eta_{0,p}^{\dag} \eta_{0,p} \ .
\eeq
so that its ground state is factorized as the product of the vacua of the pre-quench modes:
$|\psi_0 \rangle = |0_{\eta_m} \rangle \otimes |0_{\eta_0} \rangle $, where the states $|0_{\eta_{m/0}} \rangle $ satisfy $\eta_{m/0,p} |0_{\eta_{m/0}} \rangle = 0, \, \forall p$.

Here we are interested in theories that have a massive  ($m$) and a massless ($0$) mode, i.e.
in which the small momentum behavior of the two dispersions reads
\beq \label{eigenvalues}
\lambda_{m,p} = m_0, \hspace{2cm} \lambda_{0,p} = v |p| .
\eeq
For the hamiltonian \eqref{Hinitial}, considered in Ref. \cite{ruggiero2021large}, we have $m_{0}=\sqrt{\frac{gu}{K}}$
which is the mass of the massive mode, and $v=\sqrt{\frac{u_{1}u_{2}}{K_{1}K_{2}}}K$
the speed of sound of the orthogonal (massless) mode. 
As mentioned, also in the general case, we can think of the initial state as the ground state of an effective quadratic hamiltonian. In this case $m_0$ is the effective mass (whose value depends on the interaction details), while $v$ is the sound velocity of the massless mode.

\section{Preliminary results: quenches in a single CFT} \label{sec:old}

Before studying the setting of two coupled CFTs introduced above, we shorty review the results available for the simpler case of a quench in a single CFT.
In this case, the post-quench Hamiltonian is $H_{u,K}[\theta,n]$, while the initial state can have or not have a gap (i.e., an effective mass).
%can be massive, e.g. the ground state of $H_{u,K}[\theta,n]+ g/(4\pi) \int {\rm d} x \ \theta^2(x)$, or massless, namely the ground state of $H_{u_0,K_0}[\theta,n]$.

\subsection{Massive quench} \label{sec:CC}

This quench has been solved exploiting the conformal invariance of the problem,
considering an imaginary time path integral approach, that we recall in this section.
The results of this method have been developed in
~\cite{Calabrese_QuenchesCorrelationsPRL,Calabrese_QuenchesCorrelationsLong},
and later clarified and generalised in~\cite{sc-08,cardy2016further,c-14,sc-10,gc-11,dsvc-16}.
The framework is quite general and applies to quenches starting from a translationally invariant massive state $|\psi_0\rangle$, namely any state with short-range correlations.
%, such as, for example, the ground state of the gapped hamiltonian that we mentioned above.

The objects of interest are expectation values of local operators $\phi_j (x_j)$ after the quench, namely
\begin{equation} \label{correlators}
\langle \psi_0 | \phi_1 (x_1, t_1) \cdots \phi_n (x_n, t_n) | \psi_0 \rangle.
\end{equation}
In imaginary time, Eq.~\eqref{correlators} can be represented as a path integral over a strip with operator
insertions and $|\psi_0\rangle$ playing the role of boundary condition imposed at initial and final times.

A crucial point is that, exploiting the powerful tools of Renormalization Group (RG) theory of boundary
critical phenomena~\cite{diehl1997theory}, a short-ranged initial state $| \psi_0 \rangle $ can be always replaced by the appropriate RG-invariant boundary state $|B \rangle $ to which it flows.
The distance of the actual boundary state is taken into account (to leading order) introducing an extrapolation length $\tau_0$, and approximating the state as 
$|\psi_0 \rangle \simeq e^{-\tau_0 H} |B \rangle $, with $H$ the post-quench conformal hamiltonian.
Note that in terms of the creation operators $b_p^{\dagger}$, it takes the form of a squeezed state,
\begin{equation} \label{bstate}
|B\rangle \simeq \prod_{p>0} e^{b_p^\dag b^\dag_{-p}} |0_b \rangle \qquad\text{and} \qquad
|\psi_0 \rangle \simeq \prod_{p>0} e^{(1-2\tau_0 u |p| )\, b^{\dagger}_p b^{\dagger}_{-p}} |0_b \rangle,
\end{equation}
where $|0_b \rangle$ is the vacuum of bosons of the final Hamiltonian.
%{\color{blue}(incidentally, squeezed states enter as effective initial states in several different quench contexts, see e.g. \cite{stm-14,hst-16,kst-18})}.
{The extrapolation length $\tau_0$ is expected to be of the order of the inverse mass, e.g., $\tau_0 \propto 1/\sqrt{g}$ in the case of pre-quench hamiltonian \eqref{Hinitial}, and, more generally speaking, of the inverse gap for a gapped interacting theory~\cite{diehl1997theory,Calabrese_QuenchesCorrelationsLong}.}

Accordingly, Eq.~\eqref{correlators} in can be rewritten as
\begin{equation} \label{correlators2}
\langle B | \phi_1 (x_1, \tau_1) \cdots \phi_n (x_n, \tau_n) | B \rangle \, ,
\end{equation}
where the problem has been mapped to a boundary conformal field theory (BCFT).
Namely, Eq.~\eqref{correlators2} is given by a path integral over a strip of width $2\tau_0 $
with conformally invariant boundary conditions,
and operators inserted at $\tau_j$ (with $\tau_j \in [0, 2\tau_0 ]$). Eq.~\eqref{correlators2} is
often denoted as $\langle  \phi_1 (x_1, \tau_1) \cdots \phi_n (x_n, \tau_n)  \rangle_{\rm{slab} (2\tau_0)}$:
we will use this convention in Appendix~\ref{app:massive}.
One can then rely on standard CFT calculations, based on conformal maps and on the transformation of operators under those, to compute \eqref{correlators2} exactly.
The imaginary  times $\tau_j$ have to be analytically continued to $\tau_0 + i t_j $
as the final step, to recover the real time evolution.

Within this framework, very general results can be obtained for $n$-point correlation functions,
which show exponential decay in time before relaxation, with the appearance of the famous lightcone effect~\cite{Calabrese_QuenchesCorrelationsPRL,Calabrese_QuenchesCorrelationsLong,carleo2014lightcone,bonnes2014lightcone,geiger2014local,cc-05,cbp-12,dsc-17}.
The steady state shows a finite correlation length typical of a thermal system and the deviations
from a thermal state generated by the integrability of the model are small scale details not captured by the too simplistic approximation of the initial state
(the modifications necessary to observe the relaxation to a generalized Gibbs ensemble~\cite{Rigol_GGE} within this approach have been worked out in \cite{cardy2016further}).

\subsection{Massless quench} \label{sec:Cazalilla}

The path integral approach of the previous subsection does not apply to initial critical states, which are long ranged and therefore would be associated to a diverging
extrapolation length (i.e., a vanishing gap).
This case has instead been considered in Ref.~\cite{Cazalilla_MasslessQuenchLL} (see also \cite{mitra2011mode,mitra2012thermalization}, that we closely follow in terms of notation, and \cite{cazalilla2016quantum} as review on the subject), where the quench dynamics of a TLL after a sudden change of the TLL parameter, say from $K_0$ to $K_f$, is studied via a Bogoliubov approach.

In this case, the initial hamiltonian is diagonal in some operator basis $\eta_{p}$, and the final one in some other basis $b_{p}$. They are related by a Bogoliubov transformation
\begin{equation} \label{bogoliubov-cazalilla}
%\hat{\eta}_{p}\equiv
\begin{pmatrix}\eta_{p}\\
\eta_{-p}^{\dagger}
\end{pmatrix}=\begin{pmatrix}\cosh\delta & -\sinh\delta\\
-\sinh\delta & \cosh\delta
\end{pmatrix}\begin{pmatrix}b_{p}\\
b_{-p}^{\dagger}
\end{pmatrix},
%\equiv\mathfrak{B}^{-1}\hat{b}_{p},
\qquad e^{2\delta}=\frac{K_{0}}{K_{f}}.
\end{equation}
Note that this diagonalization also holds when the quench occurs in the sound velocities as well (i.e., for the more general case $ \{u_0,K_0 \} \to \{ u_f,K_f \}$). The ground state of the initial hamiltonian can be written, again, as a squeezed state in the final basis, i.e.,
\begin{equation} \label{cazalilla-state}
|0_{\eta}\rangle=\frac{1}{\textsf{N}} \prod_{p>0}e^{\textsf{W} \,  b_{p}^{\dagger}b_{-p}^{\dagger}}|0\rangle,\qquad \textsf{W} =\tanh\delta=1-2\frac{K_{f}}{K_{f}+K_{0}},
\end{equation}
where $\textsf{N}=\prod_{p>0} (1-\textsf{W}^2)^{-1/2}$ is the normalization factor.
A crucial difference as compared to the boundary state \eqref{bstate} is that here $|\textsf{W}| <1$:
this is ultimately responsible for the power-law decay of correlation functions in the state \eqref{cazalilla-state}
versus the exponential one in \eqref{bstate}.

The relaxation is towards a genuine non-equilibrium steady state, namely a
generalized Gibbs ensemble~\cite{Rigol_GGE}, determined by the underlying integrability of the model.
In fact, the late-time spatial decay is power-law and governed by an exponent that is different from
the one that governs asymptotic ground state correlations (i.e., $K_f$). In particular this Luttinger parameter gets renormalized by a function of the ratio $K_0/K_f$~\cite{mitra2011mode,mitra2012thermalization}, as one might expect from the transformation in \eqref{bogoliubov-cazalilla}.

\section{Initial state and operators' dynamics} \label{sec:ingredients}

In this section we initiate the conformal field theory study of two initially coupled TLLs in the setting of Section~\ref{sec:setting}.
We discuss first the low energy properties of the initial state, and then the operators dynamics in the Heisenberg picture.

\subsection{Features of the initial state}

\subsubsection{Leading order for small momentum}
{ A state of the form~\eqref{exactstate}, with generic ${\bf W}_p$, cannot be directly handled with conformal methods because of the non-trivial dependence of the momentum in $\textbf{W}_p$.
However, we anticipated that, invoked RG ideas, we can focus on its low energy limit, i.e., the limit $p\to0$ of $\textbf{W}_p$.
In the basis of the post-quench hamiltonian, a zero-momentum matrix with a massive and a massless mode can be parametrized as follows
\beq\label{Eq_Matrix}
W_0\equiv \textbf{W}_{p=0}=
\cos^2\varphi\ {\mathbb I}- \sin^2\varphi \ S_- ( \nu), \qquad
S_{\pm}(\nu)=
\begin{pmatrix}
\pm \cos 2 \nu & \mp \sin 2 \nu \\
\sin 2 \nu  & \cos 2 \nu
\end{pmatrix},
\eeq
where ${\mathbb I}$ is the identity matrix, and $\varphi$ and $\nu$ such that {(recall that we can always think of the state as the ground state of a quadratic hamiltonian of the form \eqref{HinitialD} and the value of the effective mass does not enter in the parameters below)}
\begin{equation} \label{varphi_nu}
\sin^2 \varphi=\frac{K_{+}}{\Gamma+K_{+}}, \qquad \nu=\text{atan}\sqrt{\frac{K_{1}}{K_{2}}}.
\end{equation}
with
\begin{equation} \label{LambdaKp}
\Gamma=
 %\frac{K_{1}K_{2}}{u_{1}u_{2}} \frac{u}{K} a =
 \frac{uK}{v}, \quad K_+ = \frac{K_1+K_2}{2}.
\end{equation}
}
The matrix $W_0$ can be diagonalized via a rotation, parametrized by an angle $\nu$
\beq \label{rotation_nu}
\begin{array}{ll}
\displaystyle b_{A,p}^\dag = \cos\nu \ b_{1,p}^\dag - \sin\nu \ b_{2,p}^\dag,
\\ \vspace{-0.2cm} \\
 \displaystyle b_{B,p}^\dag = \sin \nu \ b_{1,p}^\dag + \cos\nu \ b_{2,p}^\dag \ .
 \end{array}
\eeq

The two eigenvalues of $W_0$ are $\{1,\cos 2 \varphi \}$, associated respectively to two orthogonal sectors that we dubbed $\{ A, B \}$,
and the value of these eigenvalues determines the spectrum
and the decay of correlations of the two modes, as we saw in the previous sections and can be understood from (\ref{Eq_bb_single}). In the case of two identical systems $u_1=u_2$ and $K_1=K_2$,
the two modes are associated to the antisymmetric/symmetric fields $\theta_{\pm}$ introduced above.
However such correspondence does not hold in general.

The state~\eqref{exactstate} in the low-energy approximation is thus factorized in the basis $\{ A, B \}$.
It consists of an infinite mass state (associated to the eigenvalue $1$, see Eq.~\eqref{bstate}) in the $A$-sector,
and a massless state (associated to $|\cos 2\varphi|  <1$, see Eq.~\eqref{cazalilla-state}) in the $B$-sector.
Importantly, these two are the low energy states that characterize the dynamics studied, respectively,
by Calabrese-Cardy \cite{Calabrese_QuenchesCorrelationsPRL,Calabrese_QuenchesCorrelationsLong}, and by Cazalilla \cite{Cazalilla_MasslessQuenchLL},
as discussed in Sections \ref{sec:CC} and \ref{sec:Cazalilla}.
The dynamics in the $B$-sector can be interpreted as a quench in the TLL parameter.
Indeed, for the ground state of the hamiltonian \eqref{Hinitial}, it
corresponds to a quench   from $\Gamma$ in the initial state to $K_+$ in the post-quench hamiltonian,
as follows from identifying in Eq.~\eqref{cazalilla-state} $\tanh\delta$ with  $\cos 2\varphi = \frac{\Gamma-K_+}{\Gamma+K_+}$.

Crucially, the factorization of the state at this order is, by construction, independent of the momentum $p$, in such a way that we can define the fields
\begin{equation} \label{thetaAB_nAB}
\begin{cases}
\theta_{A}	= \frac{1}{\sqrt{K_{A}}}\left(\cos\nu\sqrt{K_{1}}\theta_{1}-\sin\nu\sqrt{K_{2}}\theta_{2}\right)\\
\theta_B	= \frac{1}{\sqrt{K_{B}}}\left(\sin\nu\sqrt{K_{1}}\theta_{1}+\cos\nu\sqrt{K_{2}}\theta_{2}\right)
\end{cases}
\begin{cases}
n_{A}=\sqrt{K_{A}}(\cos\nu\frac{n_{1}}{\sqrt{K_{1}}}-\sin\nu\frac{n_{2}}{\sqrt{K_{2}}})\\
n_{B}=\sqrt{K_{B}}(\sin\nu\frac{n_{1}}{\sqrt{K_{1}}}+\cos\nu\frac{n_{2}}{\sqrt{K_{2}}})
\end{cases}.
\end{equation}
The TLL parameters $K_A$ and $K_B$ are auxiliary variables which are free variables and, as we will see, they will not enter in the formulas for the dynamics of
physical observables.
Always to have in mind a specific example, we can plug in the above equations the value of $\nu$ in Eq. \eqref{varphi_nu}, corresponding to hamiltonian \eqref{Hinitial},
to have
\begin{equation} \label{thetaAB}
\theta_{A}	=\sqrt{\frac{K_{1}K_{2}}{2K_{+}K_{A}}}\left(\theta_{1}-\theta_{2}\right)
,\quad
\theta_B= \frac{1}{\sqrt{2K_{+}K_{B}}}\left(K_{1}\theta_{1}+K_{2}\theta_{2}\right).
\end{equation}
This equations shows that the field $\theta_{A}$ remains aligned to the massive hamiltonian term $(\theta_{1}-\theta_{2})$, cfr. Eq. \eqref{Hinitial}.

By inverting Eq.~\eqref{thetaAB_nAB} for $\{\theta_i,n_i\}$, and plugging them into the post-quench TLL hamiltonian \eqref{Hfinal}, we get
\begin{multline} \label{HLL_AB}
H =H_{u_A,K_A} [\theta_A, n_A]+ H_{u_B,K_B} [\theta_B,n_B] +\\ + \lambda_{AB}
\left(\sqrt{K_{A}K_{B}}\int dx\nabla\theta_{A}\nabla\theta_B+\frac{\pi^{2}}{\sqrt{K_{A}K_{B}}}\int dxn_{A}n_{B}\right),
%\frac{u_A}{2\pi}\int dx\left[K_{A}(\nabla\theta_{A})^{2}+\frac{\pi^{2}}{K_{A}}n_{A}^{2}\right]+\frac{u_B}{2\pi}\int dx\left[K_{B}(\nabla\theta_B)^{2}+\frac{\pi^{2}}{K_{B}}n_{B}^{2}\right]
\end{multline}
with the coupling of the $A-B$ sectors given by
\begin{equation}
\lambda_{AB}  = \frac{(u_{1}-u_{2})}{\pi}\cos\nu\sin\nu \, ,
\end{equation}
so, in general, the two sectors are coupled.
Moreover, we have
\begin{align} \label{uA}
u_A & = u_1 \cos^2\nu  + u_2 \sin^2 \nu , \\\label{uS}
u_B & =u_1 \sin^2\nu  + u_2 \cos^2 \nu ,
\end{align}
that fix the sound velocities $u_{A/B}$ of $\theta_{A/B}$.
In the case of the ground state of hamiltonian \eqref{Hinitial}, these two velocities are
$u_A= \frac{K_1 K_2}{K_+}  \frac{u}{K}$ and $u_B= \frac{1}{K_+} u K$.

We conclude this subsection with two comments. The modes $A/B$ allow us to write the initial state as a factorized squeezed state at low-energy with operators acting on
the physical vacuum of the post-quench hamiltonian. This is different from writing the state as product of the two pre-quench vacua of \eqref{HinitialD}.
As a little detour, we note that the rotation that diagonalizes $W_0$ in \eqref{Eq_Matrix} is the same one introduced in Ref. \cite{Bachas_PermeableWalls}
for permeable interfaces in CFT. Indeed, there the scattering matrix is just given by either $S_+(\nu)$ or $S_-(\nu)$. Given that $W_0$ and $S_-$ commute, they
are diagonalized by the same transformation. This observation is the starting point for a possible connection between permeable interfaces and quench problems
that will be  investigated in a forthcoming work \cite{unfolded_coupledCFT}.

\subsubsection{Beyond the leading order}

The next to leading order in $p$ of the initial state $|\psi_0\rangle$ in general breaks the factorization in the $A/B$ sectors.
It is then convenient to write these next-to-leading order terms in $p$ directly in the basis $A/B$, in which the initial state has the general form
\begin{equation} \label{stateOp}
|\psi_0\rangle = \frac{1}{\mathcal{N}} \prod_{p>0}e^{(b_{A,p}^{\dagger},b_{B,p}^{\dagger})W^{(1)}_{p} \begin{pmatrix}b_{A,-p}^{\dagger}\\
b_{B,-p}^{\dagger}
\end{pmatrix}}|0\rangle,
\qquad W_{p}^{(1)}=\begin{pmatrix}1- {2\tau_{A} u_A}|p| & 2{\gamma}|p|\\
2{\gamma}|p| & \cos2\varphi(1-{2\tau_{B} u_B}|p|)
\end{pmatrix},
\end{equation}
where the normalization $\mathcal{N}$ is reported in Appendix~\ref{app:coherent}, see Eq.~\eqref{normalization}.
Clearly $\tau_A,\tau_B$ and $\gamma$ are functions of the initial parameters $\{ K_{i},u_{i} \}$ {(e.g., for the hamiltonian~\eqref{Hinitial} they can be explicitly worked out)}, but the precise functional dependence is actually not needed. The two velocities $u_A$ and $u_B$ in $W_p$ are defined in \eqref{uA} and \eqref{uS}, respectively. %{\color{blue}$W_p^{(1)}$ is by definition $\textbf{W}_p$ up to order $O(p)$ [up to rotation R]}.

The parameter $\tau_A$ in the $AA$-component in $W_p^{(1)}$ is nothing but the extrapolation length of the massive quench introduced ad hoc in the previous section (cfr. Eq.~\eqref{bstate}). This length is interpreted as the ``distance'' from the infinite-mass state. 
%{\color{red}(here as a result of the exact calculation, in agreement with the general expectation from RG arguments in~\cite{Calabrese_QuenchesCorrelationsLong}) [erase?]}.
As already mentioned, it is expected to be of the order of the inverse gap $m_0^{-1}$.
This term is the one generating exponential decay of correlation functions.

The $O(p)$ correction to the $BB$-component (parametrised by $\tau_B)$, only produce subleading corrections, as shown in Appendix~\ref{app:coherent}.
Hence it is neglected in what follows.

The factorization of the initial state is spoiled by the presence of the off-diagonal matrix element  $\gamma \neq 0 $.
To proceed, however,  in the following we are going to assume a diagonal form of the state also at this order (i.e., $\gamma=0$).
The consequences of a non-zero value of $\gamma$ will be discussed for the correlation functions under consideration.
As we are going to argue, the role of $\gamma$ is to renormalize  subleading
power-law exponents in some cases, or just modify non-universal prefactors in other. Crucially, it will never affect the leading behaviour of the correlation functions of interest.

\subsection{Decoupling of operator dynamics}

We are going to work out the non-equilibrium dynamics in the Heisenberg picture, in which the time dependence is entirely encoded into operators,
while the state does not evolve.
A suitable rescaling of the times will allow us to always write our observables in a decoupled form with respect to the $A$ and $B$ degrees of freedom.

Let us focus on the field $\theta_1$, for $\theta_2$ the derivation is identical.
Its dynamics is given by
\begin{equation} \label{theta1t}
\theta_1 (x,t) = e^{i H_{u_1,K_1} [\theta_1,n_1] t}  \theta_1 (x) e^{-i H_{u_1,K_1} [\theta_1,n_1] t} .
\end{equation}
In the above equation we can replace
\begin{equation} \label{H1toH1H2}
i H_{u_1,K_1} [\theta_1,n_1] \, t  \to  i \left( H_{u_0,K_1} [\theta_1,n_1] +  H_{u_0,K_2} [\theta_2,n_2] \right)\, \frac{u_1}{u_0} t
\end{equation}
with $u_0$ a common (arbitrary) velocity for the two TLL hamiltonians of $\theta_{1/2}$.
The presence of $H_{u_0,K_2} [\theta_2,n_2]$ does not affect the dynamics of $\theta_1$ because the two commute.
Importantly, due to the rescaling of time, now $\theta_{1/2}$ have the same (auxiliary and fictitious) sound velocity $u_0$.
Then (cfr. Eq.~\eqref{HLL_AB})
\begin{equation} \label{H12_HAB}
H_{u_0,K_1} [\theta_1,n_1] +   H_{u_0,K_2} [\theta_2,n_2] = H_{u_0,K_A} [\theta_A,n_A] +   H_{u_0,K_B} [\theta_B,n_B]
\end{equation}
where we used that, in these rescaled time, $u_1=u_2 =u_0$ implies (see Eqs.~(\ref{uA}-\ref{uS})) $u_A=u_B = u_0 $ as well, and $\lambda_{AB}=0$ in \eqref{HLL_AB}.
This is crucial, because now the hamiltonian in the rhs of \eqref{H12_HAB} acts separately on $\theta_{A/B}$.
Finally, defining the rescaled times
\beq
t_{i}^a= t \frac{u_{i}}{u_a}, \qquad i=1,2\;\;\; {\rm and} \;\;\; a=A,B,
\eeq
the rhs of \eqref{H1toH1H2} can be recast in the form
\begin{eqnarray}
%( H_{u_0,K_A} [\theta_A,n_A] +   H_{u_0,K_B} [\theta_B,n_B]) t_1
%&=&\left( \frac{u_0}{u_A} H_{u_A,K_A} [\theta_A,n_A] +  \frac{u_0}{u_B} H_{u_B,K_B} [\theta_B,n_B]) \right)t_1 \\
 t_1^A H_{u_A,K_A} [\theta_A,n_A] + t_1^B H_{u_B,K_B} [\theta_B,n_B]
\end{eqnarray}
which plugged in \eqref{theta1t} gives
\begin{equation} \label{dyn_thetai}
\begin{split}
\theta_{1}(x,t) = \sqrt{\frac{K_A}{K_1}} \cos \nu \, \theta_A(x,t_{1}^A)+  \sqrt{\frac{K_B}{K_1}} \sin \nu \,  \theta_B(x,t_1^B), \\
\theta_{2}(x,t) =  - \sqrt{\frac{K_A}{K_2}} \sin \nu  \, \theta_A(x,t_{2}^A)+  \sqrt{\frac{K_B}{K_2}} \cos \nu \, \theta_B(x,t_{2}^B) \ ,
\end{split}
\end{equation}
where the second equation for $\theta_2$ follows from a very similar calculation.
In fact, \eqref{dyn_thetai} is nothing but the time-dependent version of Eq. \eqref{thetaAB}.

The rescaling of time introduced above is particularly important when considering observables which are functions of both $\theta_1$ and $\theta_2$, such as, for example,
of the symmetric and antisymmetric fields $\theta_{\pm}$, which decouple in terms of $\theta_{A/B}$ at any time
\begin{multline}
\label{decomposition_thetapm}
\theta_{\pm}(x,t)  =\\
\frac{1}{2}\Big[ \sqrt{K_A}   \Big( \frac{\cos \nu}{\sqrt{K_1}} \theta_A(x,t_1^A) \mp  \frac{\sin \nu}{\sqrt{K_2}}  \theta_A(x,t_2^A) \Big)
+  \sqrt{K_B} \Big( \frac{\sin \nu}{\sqrt{K_1}}  \theta_B(x,t_1^B) \pm \frac{\cos \nu}{\sqrt{K_2}} \theta_B(x,t_2^B) \Big) \Big].
%\end{array}
\end{multline}
In summary, the general idea, exploited in the following section, is to use a time rescaling to reabsorb the different velocities of the two initial
LLs into the times at which the observables are evaluated.
Hence, using rescaled modes with the same sound velocity is enough to ensure an exact decoupling into the time-dependent $\theta_A$ and $\theta_B$ at any time.
The price to pay is that equal-time observables and correlations become multi-times ones.
This is evident in \eqref{decomposition_thetapm}, where a single time in the lhs results in two different times in the rhs.

\section{Correlation functions} \label{sec:correlations}

In the previous section, we achieved the two necessary conditions to compute correlations functions, namely
\begin{itemize}
\item the factorization of the state in the basis which diagonalizes the fields $\theta_{A/B}$ (assuming to neglect the coupling $\gamma$ in Eq.~\eqref{stateOp});
\item the decoupling of the operator dynamics with respect to the same basis.
\end{itemize}
Using these properties, all the correlation functions can be computed independently in the massive and in the massless
sector as multi-point functions at different times.

In particular, for the massive sector, one can apply the method developed in \cite{Calabrese_QuenchesCorrelationsPRL,Calabrese_QuenchesCorrelationsLong},
while the computations in the massless sector are equivalent to those in \cite{Cazalilla_MasslessQuenchLL}.
%{\color{red}While it is possible to derive general results valid for an arbitrary initial state of the form \eqref{stateOp}, the formulas become soon very cumbersome.
%For this reason, we specialise this entire section to the initial state given by the ground state fo the hamiltonian \eqref{Hinitial}.
%The correlations for any other choice of the parameters $\varphi$ and $\nu$ in Eq. \eqref{Eq_Matrix} can be worked out exactly in the same manner [erase?]}
In the following, we will see how our conformal approach provides the correct result of the leading decay of the correlation functions.

\subsection{Vertex operators} \label{sub:vertex}

\subsubsection{One-point functions of $\theta_{\pm}$} \label{sub:1vertex}

As a first non-trivial example,
we consider the exponential one-point functions of $\theta_{\pm}$ (i.e., vertex operators in CFT language)
\begin{equation} \label{C1gamma}
C_{1,\gamma}^{\pm}(t)
\equiv  \left\llangle e^{i\sqrt{2}\theta_{\pm}(t)} \right\rrangle_{\gamma},
\end{equation}
where $\llangle \cdot \rrangle_{\gamma}$ denotes the expectation value over the state \eqref{stateOp}, and below we consider $\gamma=0$.
These correlations of $\theta_{\pm}$ are the experimentally relevant ones in the context of tunneled-coupled
tubes in cold atoms experiments \cite{Kuhnert_ExpEmergenceCharacteristicLength1d}. They are also the most natural also in the Hubbard and Gaudin-Yang models,
where they are associated with spin and charge sectors \cite{essler_Hubbard}.

Making use of the decomposition derived above when $\gamma=0$ (cfr. Eq.~\eqref{decomposition_thetapm}), Eq.~\eqref{C1gamma} can be cast in the following form
\begin{equation} \label{complete1vertex}
C_{1,0}^{\pm}(t)
=\left\langle e^{i\sqrt{\frac{K_{A}}{2K_{+}}}\left(\sqrt{\frac{K_{2}}{K_{1}}}\theta_{A}(t_{1})\mp\sqrt{\frac{K_{1}}{K_{2}}}\theta_{A}(t_{2})\right)}\right\rangle
\left\langle e^{i\sqrt{\frac{K_{B}}{2K_{+}}}\left(\theta_B(t_{1})\pm\theta_B(t_{2})\right)}\right\rangle,
\end{equation}
where from now on expectation values of observables which are functions of $\theta_A$ are understood to be taken on the $AA$-component of the state \eqref{stateOp}. Similarly, functions of $\theta_B$ are evaluated on the $BB$-component of \eqref{stateOp} with $\tau_B=0$ (as already mentioned, it does not contribute up to subleading corrections).
Moreover, to lighten the notation, we simply used $t_i$ instead of $t_i^{a}$, because the correlation univocally specifies whether $a=A$ or $a=B$.

Thus, we conclude that one point functions of exponential of $\theta_{\pm}$ become a product of {\it two-point} functions of $\theta_{A/B}$.
Now massive and massless part can be computed separately.

We start from the massive part. Using the approach of Section~\ref{sec:CC}, the object that we need to evaluate is a path integral over a slab of width $W= 2\tau_A$ with the two vertex operators (which are primary operators in the CFT) inserted at different points.
The final result reads (up to a unimportant prefactor)
\begin{equation} \label{massive1vertex}
\left\langle e^{i\sqrt{\frac{K_{A}}{2K_{+}}}\left(\sqrt{\frac{K_{2}}{K_{1}}}\theta_{A}(t_{1})\mp\sqrt{\frac{K_{1}}{K_{2}}}\theta_{A}(t_{2})\right)}\right\rangle
%&\sim& e^{\frac{2\pi}{W}\left(\sqrt{h_{1}h_{2}}(\pm |u_{1}+u_{2}|\mp |u_{1}-u_{2}|)-h_{1}|u_{1}|-h_{2}|u_{2}|\right)t} \\
%& \sim& e^{- \frac{\pi}{16\tau_{A} u_A }\frac{K_{1}K_{2}}{K_{+}}\left(\frac{1}{K_{1}K_{2}}(\mp |u_{1}+u_{2}| \pm|u_{1}-u_{2}|)+\frac{1}{K_{1}^{2}}|u_{1}|+\frac{1}{K_{2}^{2}}|u_{2}|\right)t} \\
\simeq  e^{- \frac{\pi}{16\tau_{A} } \frac{K}{u} \left(\frac{1}{K_{1}K_{2}}(\mp |u_{1}+u_{2}| \pm|u_{1}-u_{2}|)+\frac{1}{K_{1}^{2}}|u_{1}|+\frac{1}{K_{2}^{2}}|u_{2}|\right)t},
\end{equation}
where we used Eq.~\eqref{1pointvertex_app} in Appendix~\ref{app:massive} with the specific values $h_{1}=\frac{1}{16K_{+}}\frac{K_{2}}{K_{1}}, h_{2}=\frac{1}{16K_{+}}\frac{K_{1}}{K_{2}}$ for the conformal weights of the corresponding vertex operators.
Eq.~\eqref{massive1vertex} reproduces the leading behavior for large $t$ of $C_{1,\gamma}^{-} (t)$ obtained in Ref.~\cite{ruggiero2021large} via a Bogoliubov calculation
if we identify
\begin{equation} \label{tauAu0}
\tau_{A}=1/m_0 \, ,
%=\frac{1}{m_{0}}\frac{K_{1}K_{2}}{K_{+}K}u
\end{equation}
which is consistent with the standard interpretation of $\tau_A$ as inverse initial mass gap \cite{Calabrese_QuenchesCorrelationsPRL}.

Let us now move to the $B$ mode.
In this sector the initial state is massless, so that we can use the results in \cite{Cazalilla_MasslessQuenchLL} for the two-point function,
but generalized to the case of unequal times.
This is, once again, a standard Bogoliubov calculation (that we report in Appendix~\ref{app:massless}). The final result reads
\begin{equation} \label{massless1vertex}
\left\langle e^{i\sqrt{\frac{K_{B}}{2K_{+}}}\left(\theta_B(0,t_{1}) \pm \theta_B(0, t_{2})\right)}\right\rangle =  e^{-\frac{K_B}{4 K_+} \left\langle[\theta_B(0,t_{1})\pm\theta_B(0,t_{2})]^{2}\right\rangle}
\end{equation}
with
\begin{multline}
%\begin{align}
\label{massless1vertex_exp}
\langle[\theta_B(0,t_{1}) \pm  \theta_B(0,t_{2})]^{2}\rangle=
\frac{1}{2K_{B}}\int_{0}^{\infty}\frac{dp}{p} \times \\
\times \left\{ \mathbb{K}_{+}\left[1 \pm  \cos(p(u_{1}-u_{2})t)\right]
- \mathbb{K_{-}}\left[\frac{1}{2}(\cos(2 p u_{1} t)+\cos(2 p u_{2} t)) \pm  \cos(p(u_{1}+u_{2} t))\right] \right\}
%\end{align}
\end{multline}
and $\mathbb{K}_{\pm}=\left( \frac{\Gamma}{K_+} \pm \frac{K_+}{\Gamma}\right)$ (cfr. Eq. \eqref{KK_pm}, specialized to our quench with $K_0/K_f=\Gamma/K_+$).
According to the $\pm$ sign, this integral has or has not an infrared (small $p$) divergence. Specifically, such divergence gives $C^{+}_{1,0} (t)=0$ when plugging Eq.~\eqref{massless1vertex_exp} in \eqref{massless1vertex} (with the $+$ sign).
This in fact  is the correct result for $C^{+}_{1,\gamma} (t)$ known also from Ref. \cite{ruggiero2021large} (see Eq.~\eqref{1pointBog}  in App.~\ref{app:bogoliubov}),
which remains valid for $\gamma\neq 0$.
Conversely, Eq.~\eqref{massless1vertex} (together with \eqref{massless1vertex_exp}) leads to a power-law decay for $C_{1,0}^{-} (t)$.
The exponent is however different from the one found for $\gamma \neq 0$ (cfr. Eq. \eqref{1pointBog} in App.~\ref{app:bogoliubov}).
We will discuss this discrepancy in Section~\ref{sec:gamma}.
For now, we just point out that at leading order $\log C_{1,\gamma}^{\pm} (t) =\log C_{1,0}^{\pm} (t)$, while the same is not guaranteed for subleading corrections (the logarithm is important for the correctness of this statement, since the corrections from the massless sector to $C_{1,\gamma}^{\pm}$ are multiplicative).

\subsubsection{Two-point functions of $\theta_{\pm}$} \label{sub:2vertex}

Similar results can be found for the (exponential) two-point function of $\theta_{\pm}$
\begin{equation} \label{2point_def}
C_{2,\gamma}^{\pm}(x,t) \equiv \llangle e^{i\sqrt{2}[\theta_{\pm}(x,t)-\theta_{\pm}(0,t)]}\rrangle_{\gamma} \ .
\end{equation}
As before, we start by rewriting it in a factorized form for $\gamma=0$, i.e.
\begin{multline} \label{complete2vertex}
C_{2,0}^{\pm}(x,t)=
\left\langle e^{i\sqrt{\frac{K_{A}}{2K_{+}}} \left[\sqrt{\frac{K_{2}}{K_{1}}}\left(\theta_{A}(x,t_{1})-\theta_{A}(0,t_{1})\right)\mp\sqrt{\frac{K_{1}}{K_{2}}}\left(\theta_{A}(x,t_{2})-\theta_{A}(0,t_{2})\right)\right]}\right\rangle \times \\
\times \left\langle e^{i\sqrt{\frac{K_{B}}{2K_{+}}}\left[\theta_B(x,t_{1})-\theta_B(0,t_{1})\pm(\theta_B(x,t_{2})-\theta_B(0,t_{2}))\right]}\right\rangle.
\end{multline}
Therefore, vertex two-point functions of $\theta_{\pm}$ are mapped into the product of two four-point functions of $\theta_{A/B}$, that we can compute separately.

For the massive part, we now have a four-point function to be evaluated in the same strip geometry considered before. The result is given by (see Appendix~\ref{app:massive}, Eq.~\eqref{2pointvertex_app})
\begin{multline} \label{massive2vertex}
\left\langle e^{i\sqrt{\frac{K_{A}}{2K_{+}}}\left[\sqrt{\frac{K_{2}}{K_{1}}}\left(\theta_{A}(x,t_{1})-\theta_{A}(0,t_{1})\right)\mp\sqrt{\frac{K_{1}}{K_{2}}}\left(\theta_{A}(x,t_{2})-\theta_{A}(0,t_{2})\right)\right]}\right\rangle
\simeq \\
%- \log C_{\pm}(x,t) =
\simeq
\begin{cases}
e^{-  \frac{\pi}{16}   \frac{1}{\tau_A } \frac{K }{u}  \left[ \frac{1}{K_1^2} 2 u_1 t  +  \frac{1}{K_2^2} 2 u_2 t \mp \frac{2}{K_1K_2} ((u_1+u_2) t  - |u_1-u_2| t)   \right] }
%\qquad
&\text{$x > 2 u_1 t$}  \\
e^{- \frac{\pi}{16} \frac{1}{\tau_A } \frac{K}{u} \left[ \frac{1}{K_1^2} x +  \frac{1}{K_2^2} 2 u_2 t  \mp \frac{2}{K_1K_2} ((u_1+u_2) t - |u_1-u_2| t)   \right] }
%\qquad
&\text{$2 u_1 t > x  >   (u_1+u_2) t$} \\
e^{- \frac{\pi}{16}   \frac{1}{\tau_A } \frac{K}{u} \left[ \frac{1}{K_1^2} x +  \frac{1}{K_2^2} 2 u_2 t  \mp \frac{2}{K_1K_2} (x - |u_1-u_2| t)   \right]}
%\qquad
&\text{$ (u_1+u_2) t > x > 2 u_2 t $} \\
e^{- \frac{\pi}{16}  \frac{1}{\tau_A } \frac{K}{u} \left[ \frac{1}{K_1^2} +  \frac{1}{K_2^2}   \mp \frac{2}{K_1K_2}    \right] x \mp \frac{\pi}{16}   \frac{1}{\tau_A } \frac{K}{u}  \frac{2}{K_1K_2}  |u_1-u_2| t }
%\qquad
&\text{$ 2 u_2 t > x > |u_1-u_2| t $}\\
e^{-\frac{\pi}{16}  \frac{1}{\tau_A } \frac{K}{u} \left[ \frac{1}{K_1^2} +  \frac{1}{K_2^2}   \right] x }
%\qquad
&\text{$|u_1-u_2| t> x $}
\end{cases}
\end{multline}
where, without loss of generality, we assumed $u_1>u_2$. Again, if we fix $\tau_A$ as in \eqref{tauAu0}, this reproduces the correct exponential decay of both $C_{2,\gamma}^{\pm}$ found in~\cite{ruggiero2021large}.

Given that the $B$ sector provides algebraic correlation, the exponential contribution in Eq. \eqref{massive2vertex}
represents always the leading decay both in $x$ and $t$, as already pointed out in Ref.~\cite{ruggiero2021large}.
A possible special case is the short time regime $2u_1 t<x$  (first case in \eqref{massive2vertex}) where there is no
 $x$-dependence. Hence, the possible space dependence is entirely in the subleading power-law contributions which we now study.
The result for the $B$-part of this two-point function (derived in Appendix~\ref{app:massless}) is
\begin{multline}  \label{massless2vertex}
\langle e^{i \sqrt{\frac{K_{B}}{2K_{+}}}  \left[(\theta_B (x,t_{1})-\theta_B (0,t_{1}))\pm(\theta_B (x,t_{2})-\theta_B (0,t_{2}))\right]}\rangle \sim \\
\sim \left|x^{2}\right|^{\frac{\mathbb{K}_{+}}{8 K_{+}}}\left|1-\frac{x^{2}}{(|u_{1}-u_{2}|t)^{2}}\right|^{\pm\frac{\mathbb{K}_{+}}{8 K_{+}}}\left|1-\frac{x^{2}}{(2 u_{1} t)^{2}}\right|^{-\frac{\mathbb{K}_{-}}{16 K_{+}}}\left|1-\frac{x^{2}}{(2 u_{2} t)^{2}}\right|^{-\frac{\mathbb{K}_{-}}{16 K_{+}}}\left|1-\frac{x^{2}}{(|u_{1}+u_{2}|t)^{2}}\right|^{\mp\frac{\mathbb{K}_{-}}{8 K_{+}}}.
\end{multline}
Note that from \eqref{massless2vertex} one can easily read off all the different regimes (the same as in Eq.~\eqref{massive2vertex}), sharply separated by lightcones. Those are nonetheless smoothen out when reintroducing the ultraviolet cutoff~\cite{ruggiero2021large}.
Note also that both $K_A$ and $K_B$ cancel in the above expressions.

In the aforementioned regime of short time $( x \gg u_1t)$, Eq. \eqref{massless2vertex} gives that
$C_{2,0}^-$ is constant (the various exponents sum up to zero) and $C_{2,0}^+ \sim |x|^{{(K_0/K_f)}/{(K_+)}}= |x|^{{1}/{\Gamma}}$.
In this regime, these correlation functions match exactly the results from the Bogoliubov calculation in~\cite{ruggiero2021large} (reported, for completeness,
in Appendix~\ref{app:bogoliubov}).

In the other regimes, the power-law scaling in Eq. \eqref{massless2vertex} have exponents that are, in general,
different compared to the ones for $\gamma\neq0$.
As mentioned already for the one-point functions, this disagreement represent  the limits of the conformal method,
which does not gives access to \emph{all} power-law contributions.
Anyway, the conclusion also for the two-point function, is that the leading term is well captured is all the regimes.
We will come back to this issue in Section~\ref{sec:gamma}.

\subsection{Derivative operators} \label{sub:derivative}

We focus here on fluctuations of the initial fields ($i=1,2$)
\begin{equation}
D_{2,\gamma}^{ij}(x,t) \equiv  \llangle n_{i}(x,t)n_{j}(0,t)\rrangle_{\gamma}, \quad J_{2,\gamma}^{ij}(x,t) \equiv \llangle j_{i}(x,t)j_{j}(0,t)\rrangle_{\gamma},
\end{equation}
where $j_i (x,t)$ the current density associated to $\theta_i(x,t)$.
Density and current correlations can be related to correlators of the derivative operators, which is also a primary operator of the CFT~\cite{yellowbook}.
 In fact it holds
\begin{equation}
\begin{split}
D_{2,\gamma}^{ij}(x,t)&
=\frac{K_{i}K_{j}}{u_{i}u_{j}\pi^{2}}\left\llangle \partial_{t}\theta_{i}(x,t)\partial_{s}\theta_{j}(0,s)\right\rrangle _{\gamma}|_{s=t},\;\\
J_{2,\gamma}^{ij}(x,t)&=
\frac{1}{\pi^2} \left\llangle \partial_{x}\theta_{i}(x,t_{i})\partial_{y}\theta_{j}(y,t_{j})\right\rrangle _{\gamma} |_{y=0}.
\end{split}
\end{equation}
For definiteness, below we look at density-density correlations, while results for the currents can be similarly derived.
Following the same logic used for the vertex operators, we exploit the factorization of the state \eqref{stateOp} and the decoupling of observable to get
\begin{multline} \label{complete2derivative}
D_{2,0}^{ij}(x,t)
 =\frac{K_{i}K_{j}}{u_{i}u_{j}\pi^{2}} \times \\
\times \left((-1)^{i+j}\left({\frac{K_{1}}{K_{2}}}\right)^{\frac{(-1)^{i}+(-1)^{j}}{2}}\frac{K_{A}}{2K_{+}}\langle\partial_t\theta_{A}(x,t_{i})\partial_t \theta_{A}(0,t_{j})\rangle+\frac{K_{B}}{2K_{+}}\langle\partial_t \theta_B(x,t_{i})\partial_t \theta_B(0,t_{j})\rangle\right).
\end{multline}
We see that the first term in the above equation is associated to the massive mode,
and therefore, according to \cite{Calabrese_QuenchesCorrelationsPRL,Calabrese_QuenchesCorrelationsLong}, decays exponentially in time.
Hence, the leading term is now given by the
part associated to the massless mode, giving rise to a power
law decay, according to \cite{Cazalilla_MasslessQuenchLL}. Such power-law decay comes out very naturally within this approach.

To get this leading term, we compute
\begin{equation} \label{2derivative}
\langle\partial_t \theta_B(x,t_{i} )\partial_s \theta_B(0,s_{j})\rangle = -\frac{1}{2}\partial_{t}\partial_{s}\langle\left[\theta_B (x,t_{i})-\theta_B (0,s_{j})\right]^{2}\rangle
\end{equation}
where we defined $s_j \equiv s_j^a= ({u_j}/{u_a}) s$.
This is a two-point function at equal times when $i=j$ and at different
times when $i\neq j$. In both cases we can evaluate it using \eqref{2derivative} and the results in Appendix~\ref{app:massless}.
For $i\neq j$ we get
\begin{multline}
\langle\partial_{t}\theta_B(x,t_{i})\partial_{s}\theta_B(0,s_{j})\rangle |_{s=t} =
-\frac{u_{i}u_{j}}{4K_{+}}\times \\
\left[\frac{\mathbb{K}_{+}}{2}\left(\frac{1}{(|u_{i}-u_{j}|t+x)^{2}}+\frac{1}{(|u_{i}-u_{j}|t-x)^{2}}\right)+\frac{\mathbb{K}_{-}}{2}\left(\frac{1}{(|u_{i}+u_{j}| t+x)^{2}}+\frac{1}{(|u_{i}+u_{j}|  t-x)^{2}}\right)\right] \, ,
\end{multline}
and in the case $i=j$
\begin{align}
\langle\partial_t\theta_B(x,t_{i})\partial_s \theta_B(0,s_{i} )\rangle|_{s=t} & =-\frac{u_{i}^{2}}{4}\left(\frac{\mathbb{K}_{+}}{K_{+}}\frac{1}{x^{2}}+\frac{\mathbb{K}_{-}}{K_{+}}\frac{1}{2}\left(\frac{1}{(x-2u_{i}t)^{2}}+\frac{1}{(x+2u_{i}t)^{2}}\right)\right).
\end{align}
The density-density fluctuations finally read ($i\neq j$)
\begin{align} \label{D2_ij}
D_{2,0}^{12}(x,t)
%& \simeq-\frac{K_{1}K_{2}}{u_{1}u_{2}\pi^{2}}\frac{K_{B}}{2K_{+}}\langle\partial\theta_B(x,t_{i})\partial\theta_B(0,t_{j})\rangle\\
 & =-\frac{K_{1}K_{2}}{\pi^{2}}\frac{1}{8}
 \begin{cases}
\frac{\mathbb{K}_{+}+\mathbb{K}_{-}}{K_{+}}\frac{1}{x^{2}} & |u_{1}+u_{2}|t\ll x\\
\frac{\mathbb{K}_{+}}{K_{+}}\frac{1}{t^{2} c^{2}|u_{1}-u_{2}|^{2}}+\frac{\mathbb{K}_{-}}{K_{+}}\frac{1}{x^{2} c^{2}|u_{1}+u_{2}|^{2}} & |u_{1}-u_{2}|t\ll x\ll|u_{1}+u_{2}|t\\
\left(\frac{\mathbb{K}_{+}}{K_{+}}\frac{1}{|u_{1}-u_{2}|^{2}}+\frac{\mathbb{K}_{-}}{K_{+}}\frac{1}{|u_{1}+u_{2}|^{2}}\right)\frac{1}{t^{2}} & x\ll|u_{1}-u_{2}|t
\end{cases}
\end{align}
where above we defined $c= x/t  < \infty$, and (for $i=j$)
\begin{align} \label{D2_ii}
D_{2,0}^{ii}(x,t)
%& \simeq-\frac{K_{i}^{2}}{\pi^{2}}\frac{K_{B}}{2K_{+}}\langle\partial\theta_B(x,t_{i})\partial\theta_B(0,t_{i})\rangle\\
 & =-\frac{K_{i}^{2}}{\pi^{2}}\frac{1}{8}
 \begin{cases}
\frac{\mathbb{K}_{+}+\mathbb{K_{-}}}{K_{+}}\frac{1}{x^{2}} & 2u_{i}t\ll x\\
\frac{\mathbb{K}_{+}}{K_{+}}\frac{1}{x^{2}}+\frac{\mathbb{K}_{-}}{K_{+}}\frac{1}{(2u_{i}t)^{2}} & x\ll2u_{i}t
\end{cases} \, .
\end{align}
Similarly, for the current-current correlations we get for $i\neq j$
\begin{align} \label{J2_ij}
J_{2,0}^{12}(x,t)
%& \simeq-\frac{K_{1}K_{2}}{u_{1}u_{2}\pi^{2}}\frac{K_{B}}{2K_{+}}\langle\partial\theta_B(x,t_{i})\partial\theta_B(0,t_{j})\rangle\\
 & =-\frac{1}{8\pi^{2}}
 \begin{cases}
\frac{\mathbb{K}_{+}-\mathbb{K}_{-}}{K_{+}}\frac{1}{x^{2}} & |u_{1}+u_{2}|t\ll x\\
\frac{\mathbb{K}_{+}}{K_{+}}\frac{1}{t^{2} c^{2}|u_{1}-u_{2}|^{2}}-\frac{\mathbb{K}_{-}}{K_{+}}\frac{1}{x^{2} c^{2}|u_{1}+u_{2}|^{2}} & |u_{1}-u_{2}|t\ll x\ll|u_{1}+u_{2}|t\\
\left(\frac{\mathbb{K}_{+}}{K_{+}}\frac{1}{|u_{1}-u_{2}|^{2}}-\frac{\mathbb{K}_{-}}{K_{+}}\frac{1}{|u_{1}+u_{2}|^{2}}\right)\frac{1}{t^{2}} & x\ll|u_{1}-u_{2}|t
\end{cases}
\end{align}
and, for $i=j$,
\begin{align} \label{J2_ii}
J_{2,0}^{ii}(x,t)
%& \simeq-\frac{K_{i}^{2}}{\pi^{2}}\frac{K_{B}}{2K_{+}}\langle\partial\theta_B(x,t_{i})\partial\theta_B(0,t_{i})\rangle\\
 & =-\frac{1}{8\pi^{2}}
 \begin{cases}
\frac{\mathbb{K}_{+}-\mathbb{K_{-}}}{K_{+}}\frac{1}{x^{2}} & 2u_{i}t\ll x\\
\frac{\mathbb{K}_{+}}{K_{+}}\frac{1}{x^{2}}-\frac{\mathbb{K}_{-}}{K_{+}}\frac{1}{(2u_{i}t)^{2}} & x\ll2u_{i}t
\end{cases}.
\end{align}
As we are going to justify in the following subsection, the leading algebraic decay is not influenced by the inclusion of a $\gamma\neq 0$ term.

\subsection{Corrections from off-diagonal terms and comparison with Bogoliubov approach} \label{sec:gamma}

In the previous two subsections we calculated several correlation functions following standard RG ideas which are completely under control at equilibrium.
In particular, we focused on the first terms in a small momentum expansion of the initial state.
However, it in unclear how these RG reasonings capture the real  out-of-equilibrium time evolution of the two generally coupled TLLs.
The crucial point is represented by the generic form of the initial state in Eq. \eqref{stateOp} which shows
that at order $O(p)$ (i.e., for $\gamma\neq 0$) a term breaking the factorization of the initial state appears, modifying qualitatively
a few aspects of our approach.
It is straightforward  to realize that this term generates further algebraic decay in space and time separations, as can be understood
from the results derived in Appendix \ref{app:coherent}.
As a consequence, for all the correlation functions with a leading exponential behavior, the presence of such off-diagonal coupling only provide {\it subleading}
corrections to the result we obtained assuming factorization.
Conversely, power-law terms are in general affected by the presence of the off-diagonal term, and so are
not correctly captured within our approach working within a factorized initial state.
Interestingly (and maybe surprisingly), for all the correlations presented above, every time that the leading term is algebraic, such off-diagonal term always leave it untouched.

Let us consider as a first example $C_{2,\gamma}^{\pm} (x,t)$, defined in Eq.~\eqref{2point_def} of Section~\ref{sub:vertex}.
In this correlation, the leading behaviour with the assumption of factorized initial state is given by the exponential terms in Eq. \eqref{massive2vertex} coming from the massive sector.
The massless sector provides only a correction to the power-law multiplicative correction reported in Eq.~\eqref{massless2vertex}.
The exact result (with the correct powers), as obtained from the Bogoliubov approach, is reported in the appendix, cfr. Eq.~\eqref{eq:allregimesBog}.
It is evident that the two are in general different: in fact, while \eqref{massless2vertex} only has two free parameters (i.e., $\mathbb{K}_{\pm}$), in \eqref{eq:allregimesBog} we have four of them (i.e., $\Theta$ and $\{ a_{ij}\}$ in \eqref{eq:allregimesBog}, defined in Appendix~\ref{app:bogoliubov}). Remarkably, they exhibit the same lightcone structure.

Nonetheless, as anticipated in the previous section, it turns out that in short time regime also power laws are correctly captured.
This agreement does not come as a surprise, because, in the short-time regime (namely, before the first lightcone), the correlations reduce to the ones in the initial state. Anyhow, this obvious result comes from a non-trivial limit and was worth to test.
Moreover, since this is the only regime where power-laws become leading (cf. \eqref{massive2vertex}), the conclusion is that the leading term in both $x$ and $t$ is \emph{always} correctly captured by our approach. Note that in the intermediate regimes $x$ and $t$ have to scale in the same way (by definition, $x/t$ must be finite and within the limits defining the corresponding regime). Therefore in this case, we have effectively just one independent variable, and the leading term is exponential.
Coincidently, in the specific case $K_1=K_2$, the exponential decay in $x$ also vanishes in the prethermal regime (i.e., $2u_2 t>x >|u_1-u_2|t$), and one can verify that the power law correction is correctly described as well.

The other correlation function of interest is the density-density one that we considered in Section~\ref{sub:derivative}.
The effect of having $\gamma\neq 0$ in this correlation is to add a mixing term
$\llangle \partial \theta_A \partial \theta_B \rrangle_{\gamma}$ to Eq. \eqref{complete2derivative}.
However, it is clear that this further term decay with the \emph{same} power-law as the leading B/B term in Eq. \eqref{complete2derivative}:
therefore the presence of $\gamma$ only changes the prefactor in front of the power-law decay.

Note that there is a main difference between the two examples discussed above.
For the vertex operators, the correlations of $\theta_{A/B}$ appear in the exponent.
As a consequence, the non-diagonal contribution from $\gamma \neq 0$ are \emph{multiplicative} ones and therefore they renormalize the power-laws.
For the derivative operators, instead, the corrections in $\gamma$ are \emph{additive}, and thus they do not change the power-law exponent, but just modify the
non-universal prefactor.

\subsection{Particular limits} \label{sec:limits}

In this short subsection, we analyze how our general quench simplifies when
the velocities or the TLL parameters of the two species are the same.

\subsubsection{Same velocities $u_1=u_2$} \label{same_u}

When $u_1=u_2$ (even for $K_1\neq K_2$) in the post-quench Hamiltonian~\eqref{Hfinal} $H_{u_1,K_1}$ and $H_{u_1, K_2}$ have the same spectrum.
As a result, for any choice of the basis parametrized by $\nu$ in Eq.~(\ref{rotation_nu}), the hamiltonian corresponds to two decoupled TLLs,
as can be seen from (\ref{HLL_AB}).
The initial state then selects a particular angle $\nu$ (cf. Eq.~\eqref{varphi_nu}) that ensures the decoupling
between the massive and the massless degrees of freedom in the pre-quench Hamiltonian \eqref{Hinitial}.
In this case therefore, the initial state is {\it exactly} factorized in the two sectors and consequently the solution of non-equilibrium dynamics does not require any time rescaling
(as a trivial consequence of the equality of two velocities).
Hence, one can compute simply equal time correlation functions, and the light cone structure is largely simplified: this is evident, e.g., in \eqref{massive2vertex} where one is left with two regimes only, corresponding to a unique lightcone.

Because of  the perfect decoupling, the initial quench in the two TLLs induces a quench in the massive sector only, whereas the massless
sector remains at equilibrium in the ground state. In fact the massless quench induced in the general case $u_1 \neq u_2$ results from an effective sudden change in the TLL parameters, but in this case they turn out to be equal. Specifically, one has $\Gamma/K_+ =1$ if $u_1=u_2$ (as can be checked from Eq.~\eqref{LambdaKp}).
All previously reported results match this particular limit, but for some correlations in a singular manner requiring the restating of the short-distance cutoff.

\subsubsection{Same TLL parameters $K_1=K_2$} \label{same_K}

Contrarily to the limit of equal velocities, the one of equal TLL parameters does not bring major simplifications: the lightcone structure in \eqref{massive2vertex}, for example, remains since it is clearly related to the presence of different velocities only.

However, we note that in this case, in the correlation function \eqref{massive2vertex}, there is no exponential decay in space in the ``prethermal" regime
(namely $2u_2 t>x >|u_1-u_2|t$) for  $C_{2,0}^{+}$. Although there is still non-trivial time dependence in \eqref{massive2vertex},
the lack of exponential in $x$ can be interpreted as a prethermal ``temperature" equal to zero.
This last fact can be understood, at speculative level, by noting that in the prethermal regime the difference in the velocities is small compared to the the considered spacetime scale ($x/t \gg |u_1 -u_2|$). Therefore, as $K_1=K_2$, the total system is basically equivalent to two identical TLLs, and in that case
the symmetric mode remains in its ground state~\cite{Foini_CoupledLLsMassiveMassless}.
In this regime the power law decay in space, coming from the massless sector, which becomes dominant, is correctly
captured by our low energy approach.

\section{Conclusions} \label{conclusions}

In this work we have studied the quench dynamics of two coupled
Tomonaga-Luttinger liquids from the off-critical to the critical conformal regime.
This situation is relevant for several systems, including tunnel coupled condensates~\cite{Andrews_FirstExperimentInterferenceCondensates,Shin_ExperimentSplitCondensates1,Shin_ExperimentSplitCondensates2, Shin_ExperimentSplitCondensates3,Schumm_ExperimentMatterWaveInterferometry,Albiez_ObservationTunnellingBosonicJosephsonJunction,Gati_RealizationBosonicJosephsonJunction, Levy_ExperimentalBosonicJosephsonEffects, Kuhnert_ExpEmergenceCharacteristicLength1d}, the Hubbard model~\cite{giamarchi2004,essler_Hubbard}, the Gaudin-Yang model~\cite{yang1967some,gaudin1967systeme}
or, more generally speaking, systems with two types of degrees of freedom.

We have shown that, for what concerns the large scale properties, this non-equilibrium dynamics decouples into two independent sectors,
inducing an {\it effective} quench in each of them: one starting from a massive initial state and one from a massless one with an effective TLL parameter.
Each of them can be studied by means of the known techniques reported in Sections ~\ref{sec:CC} and ~\ref{sec:Cazalilla}.
The equal-time correlations of the coupled system map to correlation functions at different times of two uncorrelated modes.
We have also discussed that, while the leading term and the light cone structure are always well captured by our approach, this is not the case for
subleading power law corrections, generated by the coupling between these two modes (i.e., $\gamma \neq 0$ in Eq.~\eqref{stateOp}).

Moreover, a direct inspection of the correlation functions shows that while for vertex operators
the leading contribution is given by the exponential decay of the massive mode
and the massless mode acts with subleading multiplicative
power law corrections, for derivative
operators the large scale properties are determined by
the power law decay of the massless mode, while the massive part constitutes an additive correction here.
Therefore while vertex operators mimic a thermal like behavior, derivative
operators behave as at $T=0$ or more generally in a GGE~\cite{Rigol_GGE}.

A final interesting remark is that for the quench induced in the massless sector,  
the product of the speed of sound and the Luttinger parameter before and after the quench is equal (i.e., $v \Gamma = u_B K_+$ in our notation). This fact suggests that the quench respects Galilean invariance \cite{Haldane_LuttingerLiquid}.

This work paves the way to the study of quenches in systems consisting of more than a single TLL.
We conclude by providing some future possible developments in this direction.
The most natural extension of our calculation would be to apply our CFT approach to
initial states in which the orthogonal modes correspond to two massive or two massless theories.
To this aim, it is sufficient to change the parametrization of the initial squeezed state \eqref{stateOp} with two eigenvalues with absolute values equal to one (massive case)
or smaller than one (massless case).

Another  generalization would be to consider a larger number of initially coupled TLLs.
In this case, there are many different physical situations requiring different numbers of massive and massless modes.
In particular, in the case where the TLLs correspond to many different tubes,
it would be interesting to investigate how two-dimensional non-equilibrium physics emerges from the coupling of 1D systems,
as done in a very different setup in \cite{jk-15}.

Finally we mention possible connections with the works on conformal interfaces\cite{oshikawa1996defect,oshikawa1997boundary,Bachas_PermeableWalls,bachas2008fusion,bms-07}, that we plan to investigate in the future~\cite{unfolded_coupledCFT}. Our framework, in fact, can be in principle reformulated in a full path integral fashion via the \emph{unfolded picture}~\cite{oshikawa1996defect,oshikawa1997boundary}, where the initial state (living in the tensor product of two CFTs, i.e., ${\rm CFT}_1 \times {\rm CFT}_2$) is mapped to an interface (connecting two spatial regions in a single CFT, namely ${\rm CFT}_1 \bigcup \overline{\rm CFT}_2$).
%\end{itemize}

\section*{Acknowledgements}

We would like to thank J\'{e}r\^{o}me Dubail for useful discussions.
This work is supported by “Investissements d’Avenir” LabEx PALM (ANR-10-LABX-0039-PALM), EquiDystant project (LF) and by the Swiss National Science Foundation under Division II (PR and TG). PC acknowledges support from ERC under Consolidator grant number 771536 (NEMO).

% TODO: include author contributions
%\paragraph{Author contributions}
%This is optional. If desired, contributions should be succinctly described in a single short paragraph, using author initials.

% TODO: include funding information
%\paragraph{Funding information}

%\begin{appendix}

\appendix

\section{Calculations in the massive sector} \label{app:massive}

In the path integral approach in imaginary time developed in \cite{Calabrese_QuenchesCorrelationsPRL,Calabrese_QuenchesCorrelationsLong}, quantities as the one
in Eq.~\eqref{massive1vertex} are mapped to correlation functions in a strip geometry with boundary conditions corresponding to conformally invariant boundary states.
In (1+1)-dimensional BCFT, those are computed exploiting the transformations of correlation functions of (primary) operators  under conformal maps.
Let us consider, for example, two geometries in the complex plane with a boundary (say G1 and G2) with coordinate
$w$ and $z$, related by the conformal map $w(z)$.
Then the correlations of primary operators $\phi_{i}$ in the two geometries are related as
\begin{equation} \label{transformation_primaries}
\left\langle \prod_{i}\phi_{i}(w_{i},\bar{w}_{i})\right\rangle _{{\rm G1}}=\prod_{i}\left|w'(z_{i})\right|^{h_{i}}\left|\bar{w}'(\bar{z}_{i})\right|^{\bar{h}_{i}}\left\langle \prod_{i}\phi_{i}(z_{i},\bar{z}_{i})\right\rangle _{{\rm G2}},
\end{equation}
with $h_i$ and $\bar h_i$ being the holomorphic and anti-holomorphic dimensions of $\phi_i$.

\subsection{Two-point function in the slab}

To get the two-point function in \eqref{massive1vertex}, the object that we need to evaluate is a path integral over a slab (of width $W$) with the operators inserted, i.e.,
\begin{align}
 & \langle V_{\alpha_{1}}(r,\sigma_{1})V_{-\alpha_{2}}(r,\sigma_{2})\rangle_{\textrm{slab}(W)}
\end{align}
with $V_{\alpha}=e^{i\alpha\theta}$ and $\theta$ is a bosonic field. Moreover $\sigma_{i}\:(i=1,2)$ are
imaginary times, to be analitically continued to the values $\sigma_{i}\to it_{i}+W/2$
at the very end of the calculation.

Moving to complex coordinates (with points labelled by $w=r+i\sigma$), the correlation function of vertex
operators $V_{\alpha_{i}}(w_{i},\bar{w}_{i})$ on the slab geometry is first mapped by a conformal transformation to the upper-half
plane (UHP) (with coordinate $z$ s.t. ${\rm Im }(z)>0$).
The two-point function in the UHP is related to a four-point function of \emph{chiral} vertex operators
$V_{\alpha}(z_{i})$ on the complex plane, $z \in \mathbb{C}$ \cite{yellowbook}.
The details of the calculation can be found in \cite{Calabrese_QuenchesCorrelationsLong}.
The final result is
\begin{multline} \label{formulas_conformal_map1vertex}
\langle V_{\alpha_{1}}(w_{1},\bar{w}_{1})V_{-\alpha_{2}}(w_{2},\bar{w}_{2})\rangle_{\textrm{slab}(W)}
=\textrm{J}\times\langle V_{\alpha_{1}}(z_{1},\bar{z}_{1})V_{-\alpha_{2}}(z_{2},\bar{z}_{2})\rangle_{\textrm{UHP}}=\\
=\textrm{J}\times\langle V_{\alpha_{1}}(z_{1})V_{-\alpha_{2}}(z_{2})V_{-\alpha_{1}}(\bar{z}_{1})V_{\alpha_{2}}(\bar{z}_{2})\rangle_{\mathbb{C}}
%=\textrm{J}\times\left(\frac{z_{1\bar{2}}z_{2\bar{1}}}{z_{12}z_{\bar{1}\bar{2}}}\right)^{\alpha_{1}\alpha_{2}}\left(\frac{1}{z_{1\bar{1}}}\right)^{\alpha_{1}^{2}}\left(\frac{1}{z_{2\bar{2}}}\right)^{\alpha_{2}^{2}}
%=\textrm{J}\times\left(\frac{z_{1\bar{2}}z_{2\bar{1}}}{z_{12}z_{\bar{1}\bar{2}}}\right)^{2\sqrt{h_{1}h_{2}}}\left(\frac{1}{z_{1\bar{1}}}\right)^{2h_{1}}\left(\frac{1}{z_{2\bar{2}}}\right)^{2h_{2}}
%\\
=\textrm{J}\times\left(\frac{|z_{1\bar{2}}|^{2}}{|z_{12}|^{2}}\right)^{2\sqrt{h_{1}h_{2}}}\left(\frac{1}{|z_{1\bar{1}}|}\right)^{2h_{1}}\left(\frac{1}{|z_{2\bar{2}}|}\right)^{2h_{2}}
\end{multline}
conveniently expressed in terms of
\begin{align} \label{eq:geometry}
& z_{i}  = R_{i}e^{i\frac{\pi}{W}\sigma_{i}},\quad|z'_{i}|=\frac{\pi}{W}R_{i},\quad z_{i\bar{i}}=2R_{i}\sin\gamma_{i}, \nonumber \\
& |z_{ij}|^{2}  =R_{i}^{2}+R_{j}^{2}-2R_{i}R_{j}\cos(\gamma_{i}-\gamma_{j}),\\
& |z_{i\bar{j}}|^{2} =R_{i}^{2}+R_{j}^{2}-2R_{i}R_{j}\cos(\gamma_{i}+\gamma_{j}), \nonumber
\end{align}
where $\text{J}=(\pi/W)^{2(h_{1}+h_{2})}$ denotes the Jacobian factor
in \eqref{transformation_primaries}, $h_{i}=\alpha_{i}^{2}/(8 K_A)$ is the
conformal weight of the chiral operator $V_{\alpha_{i}}(z)$,
$\gamma_{i}=\pi\sigma_{i}/W$, $z_{ij} =|z_i-z_j|$, $z_{\bar{i}}=\bar{z}_i$, and $R_{i}=e^{\frac{\pi}{W}r_{i}}$ (in our case $R_1=R_2$).
 Plugging into this expression the actual coordinates, we get
\begin{multline}
\langle V_{\alpha_{1}}(w_{1},\bar{w}_{1})V_{-\alpha_{2}}(w_{2},\bar{w}_{2})\rangle_{\textrm{slab}(W)} =\\
= \left(\frac{\pi}{W}\right)^{2(h_{1}+h_{2})}\left(\frac{1-\cos\left(\frac{\pi}{W}|\sigma_{1}+\sigma_{2}|\right)}{1-\cos\left(\frac{\pi}{W}|\sigma_{1}-\sigma_{2}|\right)}\right)^{2\sqrt{h_{1}h_{2}}}\left[2\sin\left(\frac{\pi}{W}\sigma_{1}\right)\right]^{-2h_{1}}\left[2\sin\left(\frac{\pi}{W}\sigma_{2}\right)\right]^{-2h_{2}}. \label{eq:1point_differentDelta}
\end{multline}
Finally, by analytically continuing $\sigma_{i}$ to real times and taking $t_{i} \gg W$, we obtain
\begin{align} \label{1pointvertex_app}
\langle V_{\alpha_{1}}(r,t_{1})V_{-\alpha_{2}}(r,t_{2})\rangle & =\left(\frac{\pi}{W}\right)^{2(h_{1}+h_{2})}e^{\frac{2\pi}{W}\left(\sqrt{h_{1}h_{2}}(|t_{1}+t_{2}|-|t_{1}-t_{2}|)-h_{1}|t_{1}|-h_{2}|t_{2}|\right)}.
\end{align}
Upon specifying the values of $\alpha_i$ ($i=1,2$), the above equation allows to access the massive component~\eqref{massive1vertex} of the one-point function $C_{1,0}^{\pm}(t)$ (cf. Eq.~\eqref{complete1vertex} in the main text).
In particular, the difference between the symmetric and antisymmetric correlation boils down to the sign of $\alpha_2$.
It is easy to realize that this is equivalent to consider a different sign in the corresponding time $t_2$ (cf. Eq~\eqref{formulas_conformal_map1vertex}).
Collecting all these observations, the correlation of interest read
\begin{align} \label{final1pointV}
\langle V_{\alpha_{1}}(r,t_{1} )V_{-\alpha_{2}}(r, \pm t_{2})\rangle_{\textrm{slab}(W)}  \sim e^{\frac{2\pi}{W}\left(\sqrt{h_{1}h_{2}}(\pm|t_{1}+t_{2}|\mp |t_{1}-t_{2}|)-h_{1}|t_{1}|-h_{2}|t_{2}|\right)} .
%\\
% & =e^{\frac{\pi}{16\tau_{0}u_{A}}\frac{K_{1}K_{2}}{K_{+}}\left(\frac{1}{K_{1}K_{2}}(|u_{1}+u_{2}|-|u_{1}-u_{2}|)-\frac{1}{K_{1}^{2}}|u_{1}|-\frac{1}{K_{2}^{2}}|u_{2}|\right)t}
\end{align}

\subsection{Four-point function in the slab}

A similar calculation can be carried over for the the four-point function in the A-sector in Eq.~\eqref{complete2vertex}.
In the path integral formulation, the object of interest is
\begin{equation} \label{4vertex}
\langle V_{\alpha_{1}}(r,\sigma_{1})V_{\alpha_{2}}(r,\sigma_{2})V_{\alpha_{1}}(0,\sigma_{1})V_{\alpha_{2}}(0,\sigma_{2})\rangle_{\textrm{slab}(W)}
\end{equation}
where all fields and variables are defined above.
The calculation works in the exact same way, the only difference being the number of operator insertions
\begin{multline}
\langle V_{\alpha_{1}}(w_{1},\bar{w}_{1})V_{-\alpha_{2}}(w_{2},\bar{w}_{2})V_{-\alpha_{1}}(w_{3},\bar{w}_{3})V_{\alpha_{2}}(w_{4},\bar{w}_{4})\rangle_{\textrm{slab}(W)}=\\
=\textrm{J}\times\langle V_{\alpha_{1}}(z_{1},\bar{z}_{1})V_{-\alpha_{2}}(z_{2},\bar{z}_{2})V_{-\alpha_{1}}(z_{3},\bar{z}_{3})V_{\alpha_{2}}(z_{4},\bar{z}_{4})\rangle_{\textrm{UHP}}\\
%=\textrm{J}\times\left(\frac{|z_{14}||z_{23}||z_{1\bar{2}}||z_{3\bar{4}}|}{|z_{12}||z_{34}||z_{1\bar{4}}||z_{2\bar{3}}|}\right)^{\alpha_{1}\alpha_{2}}\left(\left|\frac{z_{1\bar{3}}}{z_{13}}\right|^{2}\frac{1}{z_{1\bar{1}}z_{3\bar{3}}}\right)^{\alpha_{1}^{2}}\left(\left|\frac{z_{2\bar{4}}}{z_{24}}\right|^{2}\frac{1}{z_{2\bar{2}}z_{4\bar{4}}}\right)^{\alpha_{2}^{2}}\\
=\textrm{J}\times\left(\frac{|z_{14}||z_{23}||z_{1\bar{2}}||z_{3\bar{4}}|}{|z_{12}||z_{34}||z_{1\bar{4}}||z_{2\bar{3}}|}\right)^{2\sqrt{h_{1}h_{2}}}\left(\left|\frac{z_{1\bar{3}}}{z_{13}}\right|^{2}\frac{1}{z_{1\bar{1}}z_{3\bar{3}}}\right)^{2h_{1}}\left(\left|\frac{z_{2\bar{4}}}{z_{24}}\right|^{2}\frac{1}{z_{2\bar{2}}z_{4\bar{4}}}\right)^{2h_{2}}.
\end{multline}
The relations (\ref{eq:geometry})
hold with now $i=\{1,2,3,4 \}$, and Eq.~\eqref{4vertex} corresponds to the special case $\gamma_{1}=\gamma_{3},\gamma_{2}=\gamma_{4}$,
$R_{1}=R_{2}\equiv R,R_{3}=R_{4}=1$.

Upon analytic continuation to real times, and
%$\sigma_{1}=\sigma_{3}=iu_{1}t+\tau_{0}u,\:\sigma_{2}=\sigma_{4}=-iu_{2}t+\tau_{0}u$.
taking all scales much larger than $W$, we get to the expression
\begin{multline}
\langle V_{\alpha_{1}}(r,t_{1})V_{\alpha_{2}}(r, t_{2})V_{\alpha_{1}}(0,t_{1})V_{\alpha_{2}}(0, t_{2})\rangle \\
\sim  \left[\frac{e^{\frac{2\pi}{W}\left(|t_{1}+t_{2}|\right)}}{e^{\frac{2\pi}{W}\left(|t_{1}-t_{2}|\right)}}\left(\frac{e^{\frac{2\pi}{W}r}+e^{\frac{2\pi}{W}|t_{1}-t_{2}|}}{e^{\frac{2\pi}{W}r}+e^{\frac{2\pi}{W}|t_{1}+t_{2}|}}\right)\right]^{2\sqrt{h_{1}h_{2}}}e^{-\frac{2\pi}{W}r(h_{1}+h_{2})}\left(1+\frac{e^{\frac{\pi}{W}r}}{e^{\frac{2\pi}{W}|t_{1}|}}\right)^{2h_{1}}\left(1+\frac{e^{\frac{\pi}{W}r}}{e^{\frac{2\pi}{W}|t_{2}|}}\right)^{2h_{2}}.
\end{multline}
For $C_{2,0}^{\pm}(t)$, everything simplifies to
\begin{multline} \label{2pointvertex_app}
\langle V_{\alpha_{1}}(r,t_{1})V_{\alpha_{2}}(r,\pm t_{2})V_{\alpha_{1}}(0,t_{1})V_{\alpha_{2}}(0,\pm t_{2})\rangle
%\propto & \left[\frac{e^{\frac{2\pi}{W}\left(|t_{1}\pm t_{2}|\right)}}{e^{\frac{2\pi}{W}\left(|t_{1}\mp t_{2}|\right)}}\left(\frac{e^{\frac{2\pi}{W}r}+e^{\frac{2\pi}{W}|t_{1}\mp t_{2}|}}{e^{\frac{2\pi}{W}r}+e^{\frac{2\pi}{W}|t_{1}\pm t_{2}|}}\right)\right]^{2\sqrt{h_{1}h_{2}}}\times\\
% & \qquad\qquad\times e^{-\frac{2\pi}{W}r(h_{1}+h_{2})}\left(1+\frac{e^{\frac{\pi}{W}r}}{e^{\frac{2\pi}{W}|t_{1}|}}\right)^{2h_{1}}\left(1+\frac{e^{\frac{\pi}{W}r}}{e^{\frac{2\pi}{W}|t_{2}|}}\right)^{2h_{2}}\\
\sim \\
\\
\sim \begin{cases}
e^{\frac{4\pi}{W}\left[\pm\sqrt{h_{1}h_{2}}\left(|t_{1}+t_{2}|-|t_{1}-t_{2}|\right)-h_{1}|t_{1}|-h_{2}|t_{2}|\right]} & r>2t_{1}\\
e^{\frac{4\pi}{W}\left[\pm\sqrt{h_{1}h_{2}}\left(|t_{1}+t_{2}|-|t_{1}-t_{2}|\right)\right]}e^{-\frac{2\pi}{W}r(h_{1}-2h_{2}|t_{2}|)} & |t_{1}+t_{2}|<r<2t_{1}\\
e^{\frac{4\pi}{W}\left[\pm\sqrt{h_{1}h_{2}}\left(r-|t_{1}-t_{2}|\right)\right]}e^{-\frac{2\pi}{W}r(h_{1}-2h_{2}|t_{2}|)} & 2t_{2}<r<|t_{1}+t_{2}|\\
e^{\frac{4\pi}{W}\left[\pm\sqrt{h_{1}h_{2}}\left(r-|t_{1}-t_{2}|\right)\right]}e^{-\frac{2\pi}{W}r(h_{1}+h_{2})} & |t_{1}-t_{2}|<r<2t_{2}\\
e^{-\frac{2\pi}{W}r(h_{1}+h_{2})} & r<|t_{1}-t_{2}|.
\end{cases}
\end{multline}

\section{Calculations in the massless sector} \label{app:massless}

For computing the contribution from the massless sector ($B$), we rely on the approach of \cite{Cazalilla_MasslessQuenchLL}, based on Bogoliubov transformations. In \cite{Cazalilla_MasslessQuenchLL}, the author focuses on two-point correlation functions at equal times.
When the two sound velocities are different ($u_{1}\neq u_2$), we end up in correlators at different times that we provide in what follows.

\subsection{Two-point function at different times}

Here we derive Eqs.~\eqref{massless1vertex} and \eqref{massless1vertex_exp} in the main text, that enter in the $B$-sector
contribution to Eq.~\eqref{complete1vertex}. This is just the two-point function for a quench $K_{0}\to K_{f}$ in a TLL  (below the unique sound velocity is set to $1$),
for which we use the notations introduced in Section~\ref{sec:Cazalilla}.
We consider $t_{1}\neq t_{2}$ and, without losing generality, we take $t_{1}>t_{2}$.

We compute the general correlation
\begin{equation} \label{theta_t1t2}
%C_{f}^{a}(x,t_{1};0,t_{2})=
\langle e^{i \alpha [\theta (x,t_{1})-\theta (0,t_{2})]}\rangle=e^{-\frac{\alpha^{2}}{2}\langle[\theta(x,t_{1})-\theta (0,t_{2})]^{2}\rangle} ,
%\langle[\theta_{f}(x,t_{1})\pm\theta_{f}(0,t_{2})]^{2}\rangle
\end{equation}
with $\alpha \in \mathbb{R}$ and, working in the Heisenberg picture,
the expectation value is on the ground state of the Luttinger liquid hamiltonian with Luttinger parameter $K_0$.

The correlation in the exponent in Eq.~\eqref{theta_t1t2} can be decomposed as
\begin{equation}
\langle[\theta (x,t_{1})\pm\theta (0,t_{2})]^{2}\rangle=\langle\theta (x,t_{1})^{2}\rangle+\langle\theta (0,t_{2})^{2}\rangle \pm 2 \langle\theta(x,t_{1})\theta (0,t_{2})\rangle
\label{exponentCf}
\end{equation}
where each of the terms above is a two-point function
of $\theta$ at equal or different times. Then, using the following decomposition in modes for the field (in terms of the post-quench ladder operators $b_{p}$)
%The field $\theta_f$ admits the following expansion
\begin{equation} \label{modes_thetaf}
\theta (x,t)=\frac{i}{\sqrt{L}}\sum_{p\neq0}e^{ipx}\sqrt{\frac{\pi}{2K_{f}|p|}}(b_{p}^{\dagger}(t)-b_{-p}(t)),
\end{equation}
and taking the thermodynamic limit ($L\to\infty$) we get
\begin{align}\label{eq:thetafthetaf}
\langle\theta (x,t_{1})\theta (0,t_{2})\rangle & =\frac{1}{2K_{f}}\int_{0}^{\infty}\frac{dp}{p}\cos(px)\left[\mathfrak{B}^{\dagger}U^{\dagger}(t_{1})I_{2}U(t_{2})\mathfrak{B}\right]_{22}
\end{align}
where ${\mathfrak B}$ is the Bogoliubov matrix
\begin{equation} \label{bogoliubov-cazalilla2}
%\hat{\eta}_{p}\equiv
\mathfrak{B} =
\begin{pmatrix}\cosh\delta & -\sinh\delta\\
-\sinh\delta & \cosh\delta
\end{pmatrix}, \qquad \delta=\frac{1}{2} \log(\frac{K_{0}}{K_{f}}),
\end{equation}
 and we further defined
\begin{equation}
I_{2}=\begin{pmatrix}1 & -1\\
-1 & 1
\end{pmatrix},\quad
U(t)=
\begin{pmatrix}
e^{-i|p|t} & 0\\
0 & e^{i|p|t}
\end{pmatrix}.
\end{equation}
Finally, in Eq.~\eqref{eq:thetafthetaf} we denoted as $[\cdot]_{ij}$ the elements of a given matrix.
Note that \eqref{eq:thetafthetaf} is in general not real, however we will only be interested in real combinations of terms like in Eq. \eqref{exponentCf}.
For different times ($t_{1}\neq t_{2}$) one finds
\begin{align}
\left[\mathfrak{B}^{\dagger}U^{\dagger}(t_{1})I_{2}U(t_{2})\mathfrak{B}\right]_{22} & =\left(e^{ipt_{2}}\cosh\delta-e^{-ipt_{2}}\sinh\delta\right)\left(e^{-ipt_{1}}\cosh\delta-e^{ipt_{1}}\sinh\delta\right)
\end{align}
while, at equal times ($t_{1}=t_{2}\equiv t$), it simplifies to
\begin{align} \label{matrixproductBog}
\left[\mathfrak{B}^{\dagger}U^{\dagger}(t)I_{2}U(t)\mathfrak{B}\right]_{22} & =\cosh(2\delta)-\cos(2 pt)\sinh(2\delta).
\end{align}
For $t=0$, Eq.~\eqref{matrixproductBog} simpifies to $K_{f}/K_{0}$, so that the correlations like \eqref{eq:thetafthetaf} only depends on $K_{0}$ as they should.
Eq. (\ref{exponentCf}) then reads

\begin{multline} \label{exponentCf-1}
\langle[\theta(x,t_{1})\pm\theta(0,t_{2})]^{2}\rangle
=  \frac{1}{2K_{f}}\int_{0}^{\infty}\frac{dp}{p}\left\{\mathbb{K}_{+}-\frac{\mathbb{K_{-}}}{2}(\cos(2 pt_{1})+\cos(2 pt_{2})) +  \right. \\
\left.
\pm\cos(px)\left[\mathbb{K_{+}}\cos(p|t_{1}-t_{2}|)-\mathbb{K_{-}}\cos(p|t_{1}+t_{2}|)\right] \right\},
\end{multline}
where we defined
\begin{equation} \label{KK_pm}
\mathbb{K}_{\pm} = \frac{K_0}{K_f} \pm \frac{K_f}{K_0} \, .
\end{equation}
Note that the
leading term in \eqref{exponentCf-1} diverges as $\sim1/p$, giving rise to a power decay in \eqref{theta_t1t2}.

In the case of equal spatial points, Eq.~\eqref{exponentCf-1} simplifies to
\begin{multline}
\langle[\theta (0,t_{1})\pm\theta(0,t_{2})]^{2}\rangle
=\frac{1}{2K_{f}}\int_{0}^{\infty}\frac{dp}{p} \times \\
\times \left\{\mathbb{K}_{+}\left(1\pm\cos( p(t_{1}-t_{2}))\right)-\mathbb{K_{-}}\left[\frac{1}{2}(\cos(2 pt_{1})+\cos(2 pt_{2}))\pm\cos( p(t_{1}+t_{2}))\right]\right\}.
\label{exponentCf-1-1}
\end{multline}
For $K_{0}=K_{f}$, we are computing a correlation function at
equilibrium in the ground state. Accordingly, the expression above
becomes time translational invariant (only the term involving the
times difference survives).

\subsection{Four-point function at different times}

Since the theory is quadratic, the calculation of higher point correlation functions can always be reduced to that of two-point functions. We will see it explicitly below in the case of the four-point function considered in the main text in Eq. \eqref{massless2vertex}.
We start by noting that
\begin{equation} \label{4point-cazalilla}
\langle e^{i\alpha\left[(\theta(x,t_{1})-\theta(0,t_{1}))\pm(\theta(x,t_{2})-\theta(0,t_{2}))\right]}\rangle=e^{-\frac{\alpha^{2}}{2}\langle\left[(\theta(x,t_{1})-\theta(0,t_{1}))+(\theta(x,t_{2})-\theta(0,t_{2}))\right]^{2}\rangle}
\end{equation}
which follows directly from Wick theorem. Then, we proceed by splitting the exponent in the rhs of \eqref{4point-cazalilla} in three pieces as follows
\begin{multline} \label{3terms}
\langle\left[(\theta(x,t_{1})-\theta(0,t_{1}))\pm(\theta(x,t_{2})-\theta(0,t_{2}))\right]^{2}\rangle=\\
\langle\left[\theta(x,t_{1})-\theta(0,t_{1})\right]^{2}\rangle+\langle\left[\theta(x,t_{2})-\theta(0,t_{2})\right]^{2}\rangle \\
\pm2\langle\left[\theta(x,t_{1})-\theta(0,t_{1})\right]\left[\theta(x,t_{2})-\theta(0,t_{2})\right]\rangle.
\end{multline}
This splitting is particularly convenient because each term is infared finite, so that no cutoff is needed at small $p$.

The first two terms in \eqref{3terms} are of the form (\ref{exponentCf-1}) evaluated at equal times. Performing the integral
 (with an UV cutoff)  we get
\begin{align}
\langle[\theta(x,t)-\theta(0,t)]^{2}\rangle
%& =\frac{1}{2K_{f}}\int_{0}^{\infty}\frac{dp}{p}\left[\mathbb{K}_{+}-\mathbb{K_{-}}\cos(2t)-\cos(px)\left[\mathbb{K_{+}}-\mathbb{K_{-}}\cos(2t)\right]\right]\label{exponentCf-1-2}\\
% & =\frac{\mathbb{K}_{+}}{2K_{f}}\int_{0}^{\infty}\frac{dp}{p}\left[1-\cos(px)\right]-\frac{\mathbb{K}_{-}}{2K_{f}}\int_{0}^{\infty}\frac{dp}{p}\cos(2t)\left[1-\cos(px)\right]\\
 & =\frac{\mathbb{K}_{+}}{2K_{f}}\frac{1}{2}\log|x^{2}|-\frac{\mathbb{K}_{-}}{2K_{f}}\frac{1}{2}\log\left|1-\frac{x^{2}}{(2t)^{2}}\right|.
\end{align}
For the last term in \eqref{3terms}, we find
\begin{multline}
2\langle\left[\theta(x,t_{1})-\theta(0,t_{1})\right]\left[\theta(x,t_{2})-\theta(0,t_{2})\right]\rangle=\\
=\frac{1}{K_{f}}\int_{0}^{\infty}\frac{dp}{p}\left(1-\cos(px)\right)\left[\mathbb{K_{+}}\cos(t_{1}-t_{2})-\mathbb{K_{-}}\cos(t_{1}+t_{2})\right]\\
= \left( \frac{\mathbb{K}_{+}}{K_{f}}\frac{1}{2}\log\left|1-\frac{x^{2}}{(t_{1}-t_{2})^{2}}\right|-\frac{\mathbb{K}_{-}}{K_{f}}\frac{1}{2}\log\left|1-\frac{x^{2}}{(t_{1}+t_{2})^{2}}\right| \right).
\end{multline}
Putting everything together and performing trivial algebraic simplifications, we get
\begin{multline}
\langle\left[(\theta (x,t_{1})-\theta(0,t_{1}))\pm(\theta (x,t_{2})-\theta(0,t_{2}))\right]^{2}\rangle=\\
\log\left|x^{2}\right|^{\frac{\mathbb{K}_{+}}{2K_{f}}}\left|1-\frac{x^{2}}{(t_{1}-t_{2})^{2}}\right|^{\pm\frac{\mathbb{K}_{+}}{2K_{f}}}\left|1-\frac{x^{2}}{(2t_{1})^{2}}\right|^{-\frac{\mathbb{K}_{-}}{4K_{f}}}\left|1-\frac{x^{2}}{(2t_{2})^{2}}\right|^{-\frac{\mathbb{K}_{-}}{4K_{f}}}\left|1-\frac{x^{2}}{(t_{1}+t_{2})^{2}}\right|^{\mp\frac{\mathbb{K}_{-}}{2K_{f}}} \, .\label{eq:allregimesCaz}
\end{multline}

\section{Calculations in the exact state: Bogoliubov approach} \label{app:bogoliubov}

For comparison, we briefly sketch the calculations for the same correlation functions within the Bogoliubov approach.
More details can be found in Ref. \cite{ruggiero2021large}.

\subsection{One-point function of $\theta_{\pm}$}

The one-point function in Eq. \eqref{C1gamma} can be written as
\begin{equation} \label{1pointBog}
C_{1,\gamma}^{\pm} (t)=e^{-\llangle\theta_{\pm}^{2}(t)\rrangle_{\gamma}}=e^{-\int dp\llangle\theta_{\pm,p}(t)\theta_{\pm,-p}(t)\rrangle_{\gamma}},
\end{equation}
with
\begin{equation}\label{eq:sum_ij}
\llangle\theta_{\pm,p}(t)\theta_{\pm, -p}(t)\rrangle_{\gamma}=\frac{1}{2}\sum_{i,j=1}^2(-1)^{i+j}\llangle\theta_{i,p}(t)\theta_{j,-p}(t)\rrangle_{\gamma},
\end{equation}
where used the decomposition
\begin{equation} \label{decomposition_app}
\theta_{i}(x,t)=\sum_{p}e^{-ipx}\theta_{i,p}(t),
\end{equation}
and ($\alpha_{i,p}=\frac{\pi}{K_i |p|}$)
\begin{equation}
\theta_{i,p}(t)=\cos(u_{i}|p|t)\theta_{i,p}(0)-\alpha_{i,p}\sin(u_{i}|p|t)n_{i,p}(0).
\end{equation}
Hence, the expectation value in the exponent in \eqref{1pointBog} is
\begin{multline} \label{integrand}
\llangle\theta_{i,p}(t)\theta_{j,-p}(t)\rrangle_{\gamma}=\\
=\cos(u_{i}pt)\cos(u_{j}pt)\llangle\theta_{i,p}\theta_{j,-p}\rrangle_{\gamma}+\sin(u_{i}pt)\sin(u_{j}pt)\alpha_{i,p}\alpha_{j,-p}\llangle n_{i,p} n_{j,-p}\rrangle_{\gamma}.
\end{multline}
The small $p$ expansion of \eqref{integrand} reads
\begin{equation}
\llangle\theta_{i,p}(t)\theta_{j,-p}(t)\rrangle_{\gamma}= \frac{\mathcal{A}_{ij} (p)}{p^2}+\frac{\mathcal{B}_{ij} (p)}{p}+O(0),
\label{exptt}
\end{equation}
with $\mathcal{A}_{ij} (p)$ and $\mathcal{B}_{ij} (p)$ regular for $p \to 0$.
The leading contribution in \eqref{1pointBog} comes from the term $\propto1/p^{2}$.
%(specifically, from the term containing the initial density fluctuations).
This contribution was explicitly computed in \cite{ruggiero2021large} and gives an exponential decay in \eqref{1pointBog}.

Now, we consider the next-to-leading contribution $\propto1/p$ in \eqref{exptt}.
The explicit expression for $\mathcal{B}$ is
\begin{equation} \label{2point_p}
\mathcal{B}_{ij} (p)=\Theta_{ij}\cos(u_{i}pt)\cos(u_{j}pt)+\Pi_{ij}\sin(u_{i}pt)\sin(u_{j}pt)
\end{equation}
where we defined
\begin{equation} \label{Theta}
\Theta_{ij}\equiv\lim_{p\to0}p\llangle\theta_{i,p}\theta_{j,-p}\rrangle_{\gamma} = \frac{\pi}{4\Gamma}\equiv \Theta \ ,
%\qquad \Pi_{ij}=\lim_{p\to0}p\:\alpha_{i,p}\alpha_{j,-p}\llangle n_{i,p}(t)n_{j,-p}\rrangle_{\gamma}.
\end{equation}
and
\begin{equation} \label{Pi}
\Pi_{ij}=\lim_{p\to0}p\:\alpha_{i,p}\alpha_{j,-p}\llangle n_{i,p}(t)n_{j,-p}\rrangle_{\gamma} = \Theta \, a_{ij}^2.
\end{equation}
Integration over momentum of Eq.~\eqref{2point_p}  gives ($y=pt$)
\begin{equation} \label{Bij}
\int \frac{dp}{2\pi}\: \frac{\mathcal{B}_{ij} (p)}{p}  =  \frac{\Theta}{2\pi} \left\{ \frac{1+a^2_{ij}}{2}\int dy\frac{\cos(|u_{i}-u_{j}|y)}{y}+\frac{1-a^2_{ij}}{2}\int dy\frac{\cos(|u_{i}+u_{j}|y)}{y} \right\} .
\end{equation}
%
%Let us assume that the velocities inside the cosines only gives subleading corrections we are not interested in.
Then, using Eq.~\eqref{eq:sum_ij}, and the expansion
\begin{equation}
\llangle\theta_{\pm}(t)\theta_{\pm}(t)\rrangle_{\gamma}  = \int \frac{dp}{2\pi}\: \left( \frac{\mathcal{A}_{\pm} (p)}{p^2}+\frac{\mathcal{B}_{\pm} (p)}{p}+O(0) \right),
\label{expttpm}
\end{equation}
we find
\begin{equation}
\int \frac{dp}{2\pi}\: \frac{\mathcal{B}_{\pm} (p)}{p}
% \sim\frac{1}{2\pi}\sum(\pm)^{i+j}\langle\theta_{i}(t)\theta_{j}(t)\rangle\label{eq:sum_ij-1-1-1-1}\\
%\sim \frac{(A+B)_{11}+(A+B)_{22}}{2}\int\frac{dy}{y}  +\\
%+\left[\frac{(A-B)_{11}}{2}+\frac{(A-B)_{22}}{2}\pm {(A-B)_{12}}\pm {(A+B)_{12}}\right]\int\frac{dy}{y}\cos y\\
=\frac{\Theta}{2\pi} \left\{\frac{a^2_{11}+a^2_{22}}{2}\:\int\frac{dz}{z}(1-\cos z) +\int\frac{dz}{z}(1+(1\pm2)\cos z) \right\}
\end{equation}
The above integral is convergent in one case ($-$), while diverges in the other ($+$) ones (due to the infrared behavior).
Since it appear in the exponent for $C_{1,\gamma}^{\pm}$ (cfr. \eqref{1pointBog}), it implies an algebraic decay at large $t$ for
that $C_{1,\gamma}^{-}$, and gives $C_{1,\gamma}^+=0$.

\subsection{Two-point function of $\theta_{\pm}$}

We can similarly derive the two-point function \eqref{2point_def}, i.e.
\begin{equation} \label{2point_exp}
C_{2,\gamma}^{\pm} (x, t) = e^{- \llangle  [\theta_{\pm}(x,t)-\theta_{\pm}(0,t)]^2 \rrangle_{\gamma} } \ .
\end{equation}
Let us start by reintroducing the space dependence in \eqref{expttpm} as follows
\begin{align} \label{expttpmx}
\llangle\theta_{\pm}(x,t)\theta_{\pm}(y,t)\rrangle_{\gamma} = \int_{0}^{\infty}\frac{dp}{\pi}e^{ip(x-y)}  \left( \frac{\mathcal{A}_{\pm} (p)}{p^2}+\frac{\mathcal{B}_{\pm} (p)}{p}+O(0) \right),
\end{align}
We are interested in the term in \eqref{expttpmx} whose integrand is $\propto 1/p$. Using \eqref{Bij}, the latter reads
\begin{multline}
\int_{0}^{\infty}\frac{dp}{\pi}e^{ip(x-y)} \frac{\mathcal{B}_{\pm} (p)}{p}
 = \frac{\Theta}{2\pi}\Big[\frac{(1+a_{11}^{2})+(1+a_{22}^{2})}{2}\int\frac{dp}{p}e^{ip(x-y)} +\\
+\frac{(1-a_{11}^{2})}{2}\int\frac{dp}{p}e^{ip(x-y)}\cos(2u_{1}pt)+\frac{(1-a_{22}^{2})}{2}\int\frac{dp}{p}e^{ip(x-y)}\cos(2u_{1}pt)\\
\quad\pm(1+a_{12}^{2})\int\frac{dp}{p}e^{ip(x-y)}\cos(|u_{1}-u_{2}|pt)\pm(1-a_{12}^{2})\int\frac{dp}{p}e^{ip(x-y)}\cos(|u_{1}+u_{2}|pt)\Big]
\end{multline}
 Finally, from the above expression, we get for $ \llangle\left[\theta_{\pm}(x,t)-\theta_{\pm}(0,t)\right]^{2}\rrangle_{\gamma}$ a contribution of the form
\begin{equation} \label{eq:allregimesBog}
% \llangle\left[\theta_{\pm}(x,t)-\theta_{\pm}(0,t)\right]^{2}\rrangle_{\gamma}=\\
\frac{\Theta}{2\pi}\log|x^{2}|^{\frac{(2+a_{11}^{2}+a_{22}^{2})}{2}}\left|1-\frac{x^{2}}{(2u_{1}t)^{2}}\right|^{\frac{(1-a_{11}^{2})}{2}}\left|1-\frac{x^{2}}{(2u_{2}t)^{2}}\right|^{\frac{(1-a_{22}^{2})}{2}}\left|1-\frac{x^{2}}{(|u_{1}-u_{2}|t)^{2}}\right|^{\pm(1+a_{12}^{2})}\left|1-\frac{x^{2}}{(|u_{1}+u_{2}|t)^{2}}\right|^{\pm(1-a_{12}^{2})}.
\end{equation}
Plugged in \eqref{2point_exp}, this is the final result.

\section{Calculations in the exact state: Coherent states} \label{app:coherent}

Some of the calculations in the main text are more easily done in
the coherent states basis (in a path integral fashion), that we now review. To begin with, we consider
a simple squeezed state of the form
\begin{equation}
|\psi\rangle=\prod_{p>0}e^{\textsf{W}_{p}b_{p}^{\dagger}b_{-p}^{\dagger}}|0\rangle,
\end{equation}
with $\textsf{W}_p \in \mathbb{C}$. Let us define the coherent states $|z_p\rangle$  as follows
\begin{align} \label{coherent_states_def1}
b_{p}|z_{p}\rangle=z_{p}|z_{p}\rangle, & \quad|z_{p}\rangle=e^{z_{p}b_{p}^{\dagger}-z_{p}^{*}b_{p}}|0\rangle,\\
\langle z_{p}|w_{p}\rangle=e^{-\frac{1}{2}\left(|z_{p}|^{2}+|w_{p}|^{2}-2z_{p}^{*}w_{p}\right)}, & \quad\mathbb{I}=\int\prod_{p}\frac{dz_{p}d\bar{z}_{p}}{\pi}|z\rangle\langle z|
\end{align}
where $b_{p}^{(\dagger)}$ are operators, $z_{p},w_{p}\in\mathbb{C}$,
and $|z\rangle=\otimes_{p}|z_{p}\rangle$. The norm of $|\psi\rangle$
\begin{equation}
\langle\psi|\psi\rangle=\int\prod_{p}\frac{dz_{p}d\bar{z}_{p}}{\pi}\langle\psi|z\rangle\langle z|\psi\rangle,
\end{equation}
is computed as follows.
Using the definitions in \eqref{coherent_states_def1}, we find
\begin{align}
\langle\psi|z\rangle & =\prod_{p>0}\langle0|e^{\textsf{W}_{p}b_{p}b_{-p}}|z_{p},z_{-p}\rangle=\prod_{p>0}e^{\textsf{W}_{p}z_{p}z_{-p}}e^{-\frac{1}{2}(|z_{p}|^{2}+|z_{-p}|^{2})}\,.
\end{align}
Moreover, using also that $\langle z|\psi\rangle=\langle\psi|z\rangle^{*}$,
the norm of $|\psi\rangle$ is written as a gaussian integral, which
can be computed explictly
\begin{align} \label{normalization0}
\langle\psi|\psi\rangle & =\prod_{p>0}\frac{1}{1-\textsf{W}_{p}^{2}}\equiv \textsf{N}^{2}\:.
\end{align}
Moving to correlation functions, and taking into account the above
normalization, we similarly find
\begin{equation}\label{Eq_bb_single}
\langle\frac{\psi}{\textsf{N}}|b_{q}b_{q}^{\dagger}|\frac{\psi}{\textsf{N}}\rangle=\frac{1}{1-\textsf{W}_{q}^{2}},\quad\langle\frac{\psi}{\textsf{N}}|b_{q}^{\dagger}b_{q}|\frac{\psi}{\textsf{N}}\rangle=\frac{\textsf{W}_{q}^{2}}{1-\textsf{W}_{q}^{2}}\:,
\end{equation}
where we used the commutation relations to make $b_{p}$ act on $|z\rangle$.

We then consider the squeezed state of interest in this work, namely
of the form
\begin{equation} \label{state_generic}
|\psi\rangle=\prod_{p>0}e^{(b_{A,p}^{\dagger},b_{B,p}^{\dagger})\mathbb{W}_{p}\begin{pmatrix}b_{A,-p}^{\dagger}\\
b_{B,-p}^{\dagger}
\end{pmatrix}}|0\rangle,\qquad \mathbb{W}_{p}=\begin{pmatrix}w_{AA} & w_{AB}\\
w_{AB} & w_{BB}
\end{pmatrix}
\end{equation}
where in $w_{ab}=w_{ab}^p$ ($a,b \in \{ A,S\}$) the $p$-dependence is implicit, and, for simplicity, we assumed them to be real.
First we (re)define the coherent states as
\begin{equation}
|z\rangle=\bigotimes_{p} |z_{A,p} \rangle \otimes | z_{B,p}\rangle,\quad b_{A/B,p}|z_{A/B,p}\rangle=z_{A/B,p}|z_{A/B,p}\rangle\,.
\end{equation}
and we start again from computing the norm of $|\psi\rangle$. Repeating
the same steps above, we find
\begin{align}
\langle\psi|z\rangle & =\prod_{p>0}e^{(z_{A,p},z_{B,p})\mathbb{W}_{p}\begin{pmatrix}z_{A,-p}\\
z_{B,-p}
\end{pmatrix}}e^{-\frac{1}{2}(|z_{A,p}|^{2}+|z_{A,-p}|^{2}+|z_{B,p}|^{2}+|z_{B,-p}|^{2})}\:,
\end{align}
and $\langle z|\psi\rangle=\langle\psi|z\rangle^{*}$. Using this,
the norm of the state \eqref{state_generic} can be put in the form
\begin{equation}
\prod_{p>0}\int\frac{d\hat{Z}_p}{\pi^{4}}e^{\hat{Z}_p^{T}\mathbb{\hat{\mathbb{M}}}_p\hat{Z}_p},\quad\hat{\mathbb{M}}_p=\begin{pmatrix}\begin{pmatrix} & \mathbb{W}_p\\
\mathbb{W}_p
\end{pmatrix} & -\mathbb{I}_{4}\\
-\mathbb{I}_{4} & \begin{pmatrix} & \mathbb{W}_p\\
\mathbb{W}_p
\end{pmatrix}
\end{pmatrix}
\end{equation}
where we defined the vector $\hat{Z}_p=(z_{A,p},z_{B,p},z_{A,-p},z_{B,-p},z_{A,p}^{*},z_{B,p}^{*},z_{A,-p}^{*},z_{B,-p}^{*})$, $\mathbb{I}_{4}$ is a $4\times4$ identity matrix, and $\hat{\mathbb{M}}_p$
results in a $8\times8$ symmetric matrix. Expoiting its gaussian
nature, the above integral can be evaluated analytically to get
\begin{align} \label{normalization}
\mathcal{N}^{2} & \equiv\langle\psi|\psi\rangle=
%\frac{1}{\pi^{4}}\int d^{8}\hat{Z}_{p}e^{-\hat{Z}_{p}^{T}\mathbb{\hat{\mathbb{M}}}_{p}\hat{Z}_{p}}=
\prod_{p>0}\frac{1}{\sqrt{\det\hat{\mathbb{M}}_{p}}}\nonumber \\
 & =\prod_{p>0}\left[\left(w_{AA}+w_{BB}+w_{AA}w_{BB}+w_{AB}^{2}-1\right)\left(w_{AA}+w_{BB}+w_{AA}w_{BB}-w_{AB}^{2}+1\right)\right]^{-1}.
\end{align}
Similarly, correlation functions can be evaluated making use of the property
of gaussian integral
\begin{equation}
\int d\hat{Z}e^{\hat{Z}^{T}\mathbb{\hat{\mathbb{M}}}\hat{Z}}f(\hat{Z})=\sqrt{\frac{\pi^{n}}{\det\hat{\mathbb{M}}_{p}}}\left(e^{-\sum_{ij}\mathbb{\hat{\mathbb{M}}}_{ij}^{-1}\partial_{i}\partial_{j}}\right)f(\hat{Z})|_{\hat{Z}=0}.
\end{equation}
For example, with the definitions above,
\begin{multline}
\langle\psi|b_{A,-q}b_{A,q}^{\dagger}|\psi\rangle =\\
\prod_{p\neq q}\frac{1}{\sqrt{\det\hat{\mathbb{M}}_{p}}}\frac{1}{\sqrt{\det\hat{\mathbb{M}}_{q}}}\left(e^{-\sum_{ij}\mathbb{\hat{\mathbb{M}}}_{ij}^{-1}\partial_{i}\partial_{j}}\right)\hat{Z}_{q,1}\hat{Z}_{q,3}|_{\hat{Z}=0}=\prod_{p\neq q}\frac{1}{\sqrt{\det\hat{\mathbb{M}}_{p}}}\left(\frac{-2\hat{\mathbb{M}}_{q,13}^{-1}}{\sqrt{\det\hat{\mathbb{M}}_{q}}}\right).
\end{multline}
All the two-point functions of $b_{A/B,p}^{(\dagger)}$ can be collected in the following $4\times4$ matrix
\begin{align}
\mathbb{B}{}_{p} & =\langle\begin{pmatrix}b_{A,p}^{\dagger}\\
b_{A,-p}\\
b_{B,p}^{\dagger}\\
b_{B,-p}
\end{pmatrix}\begin{pmatrix}b_{A,p} & b_{A,-p}^{\dagger} & b_{B,p} & b_{B,-p}^{\dagger}\end{pmatrix}\rangle= \begin{pmatrix}m_{15}-1 & m_{13} & m_{25} & m_{23}\\
m_{13} & m_{15} & m_{23} & m_{25}\\
m_{25} & m_{23} & m_{26}-1 & m_{24}\\
m_{23} & m_{25} & m_{24} & m_{26}
\end{pmatrix}
%\\
% & =\left\langle \begin{pmatrix}(b_{A,p}b_{A,p}^{\dagger}-1) & b_{A,p}^{\dagger}b_{A,-p}^{\dagger} & b_{B,p}b_{A,p}^{\dagger} & b_{A,p}^{\dagger}b_{B,-p}^{\dagger}\\
%b_{A,-p}b_{A,p} & b_{A,-p}b_{A,-p}^{\dagger} & b_{A,-p}b_{B,p} & b_{A,-p}b_{B,-p}^{\dagger}\\
%b_{A,p}b_{B,p}^{\dagger} & b_{B,p}^{\dagger}b_{A,-p}^{\dagger} & (b_{B,p}b_{B,p}^{\dagger}-1) & b_{B,p}^{\dagger}b_{B,-p}^{\dagger}\\
%b_{B,-p}b_{A,p} & b_{B,-p}b_{A,-p}^{\dagger} & b_{B,-p}b_{B,p} & b_{B,-p}b_{B,-p}^{\dagger}
%\end{pmatrix}\right\rangle
\end{align}
where $m_{kl}= -2 \hat{\mathbb{M}}_{p,kl}^{-1}$,
and we exploited the symmetries of $\mathbb{B}_{p}$. Expectation
values are understood on the normalized state $|\psi\rangle/\mathcal{N}$.
In particular, we want to consider the $O(p^{0})$ of $\mathbb{B}_{p}$, for the
squeezed state in \eqref{state_generic} with $\mathbb{W}_{p}=W_{p}$
(cfr. Eq.~\eqref{stateOp} in the main text), so that expectation values are given by $\llangle \cdot \rrangle_{\gamma}$.

Using the following definitions (analogous to \eqref{Theta} and \eqref{Pi})
\begin{equation} \label{ThetaPiAB}
\Theta_{a,b}\equiv\lim_{p\to0}p\llangle\theta_{a,p}\theta_{b,-p}\rrangle_{\gamma},\qquad \Pi_{a,b}=\lim_{p\to0}p\:\alpha_{a,p}\alpha_{b,-p}\llangle n_{a,p}(t)n_{b,-p}\rrangle_{\gamma}, \quad a,b \in \{ A,B \},
\end{equation}
and by expanding $\theta_{A/B,p}$ and $n_{A/B,p}$ in terms of $b^{\dagger}_{A/B,p}$, one can check that
\begin{equation} \label{ThetaAB}
\Theta_{AA}= \Theta_{AB}=0,\quad \Theta_{BB}= \frac{\pi}{2 K_B} \frac{\cos 2\varphi-1}{\cos2\varphi+1}
%\left(=\frac{K_{f}}{K_{0}}\right)
\end{equation}
%\begin{align}
%\langle\theta_{A,p}\theta_{B,p}\rangle & =\langle\left(b_{A,p}^{\dagger}-b_{A,-p}\right)\left(b_{B,-p}^{\dagger}-b_{B,p}\right)\rangle=0\label{eq:A_AB}\\
%\langle\theta_{A,p}\theta_{A,p}\rangle & =\langle\left(b_{A,p}^{\dagger}-b_{A,-p}\right)\left(b_{A,-p}^{\dagger}-b_{A,p}\right)\rangle=0\\
%\langle\theta_{B,p}\theta_{B,p}\rangle & =\langle\left(b_{B,p}^{\dagger}-b_{B,-p}\right)\left(b_{B,-p}^{\dagger}-b_{B,p}\right)\rangle=\frac{2}{1+\cos2\varphi}-1\left(=\frac{K_{f}}{K_{0}}\right)
%\end{align}
namely they are independent on the value of $\gamma$.
%So the intuition by Pasquale that at
%$t=0$ things should not depend on $p$ seems to be correct. This
%would not be the case for $\langle n_{A}n_{B}\rangle$, but in this
%case this will affect the amplitude only, which is known to be non-universal.
This is not the case for $\Pi_{ab}$, in which case one finds
\begin{equation} \label{PiAB}
\begin{split}
\Pi_{AA}=  \frac{\pi}{2K_A} \left( \frac{\gamma^{2}}{u_A^2\tau_{A}^{2}} \frac{2}{\cos2\varphi-1}-1 \right),\; \\
\Pi_{AB}= \frac{\pi}{2\sqrt{K_A K_B}} \left( \frac{\gamma}{u_A \tau_{A}} \frac{2}{\cos2\varphi -1} \right) ,\; \\
\Pi_{BB}= \frac{\pi}{2 K_B} \left( \frac{1+\cos 2\varphi}{1-\cos2\varphi} \right) .
\end{split}
\end{equation}
Finally, note that $\tau_{B}$ never enters in the above expressions.

\nolinenumbers


\begin{thebibliography}{10}
\providecommand{\url}[1]{\texttt{#1}}
\providecommand{\urlprefix}{URL }
\expandafter\ifx\csname urlstyle\endcsname\relax
  \providecommand{\doi}[1]{doi:\discretionary{}{}{}#1}\else
  \providecommand{\doi}{doi:\discretionary{}{}{}\begingroup
  \urlstyle{rm}\Url}\fi
\providecommand{\eprint}[2][]{\url{#2}}

\bibitem{Bloch_RevUltracold}
I.~Bloch, J.~Dalibard and W.~Zwerger,
\newblock \emph{Many-body physics with ultracold gases},
\newblock Rev. Mod. Phys. \textbf{80}, 885 (2008) \doi{10.1103/RevModPhys.80.885}.

\bibitem{Langen_RevUltracoldAtoms}
T.~Langen, R.~Geiger and J.~Schmiedmayer,
\newblock \emph{Ultracold atoms out of equilibrium},
\newblock Ann. Rev. Cond. Matter Phys. \textbf{6}, 201
  (2015),
\newblock \doi{10.1146/annurev-conmatphys-031214-014548}.

\bibitem{TEBD}
G. Vidal, \textit{Efficient simulation of one-dimensional quantum many-body systems},
{Phys. Rev. Lett. {\bf 93}, 040502 (2004)}, \doi{10.1103/PhysRevLett.93.040502}

\bibitem{white1992density}
S.~R. White,
\newblock \emph{Density matrix formulation for quantum renormalization groups},
\newblock Phys. Rev. Lett. \textbf{69}, 2863 (1992),
\newblock \doi{10.1103/PhysRevLett.69.2863}.

\bibitem{schollwock2005density}
U.~Schollw\"ock,
\newblock \emph{The density-matrix renormalization group},
\newblock Rev. Mod. Phys. \textbf{77}, 259 (2005),
\newblock \doi{10.1103/RevModPhys.77.259}.

\bibitem{daley2004time}
A.~J. Daley, C.~Kollath, U.~Schollwöck and G.~Vidal,
\newblock \emph{Time-dependent density-matrix renormalization-group using
  adaptive effective Hilbert spaces},
\newblock J. Stat. Mech.  P04005 (2004),
\newblock \doi{10.1088/1742-5468/2004/04/p04005}.

\bibitem{white2004real}
S.~R. White and A.~E. Feiguin,
\newblock \emph{Real-time evolution using the density matrix renormalization group},
\newblock Phys. Rev. Lett. \textbf{93}, 076401 (2004),
\newblock \doi{10.1103/PhysRevLett.93.076401}.

\bibitem{giamarchi2004}
T.~Giamarchi,
\newblock \emph{Quantum physics in one dimension},
\newblock Oxford University Press (2003), \doi{10.1093/acprof:oso/9780198525004.001.0001}

\bibitem{cazalilla2016quantum}
M.~A. Cazalilla and M.-C. Chung,
\newblock \emph{Quantum quenches in the Luttinger model and its close relatives},
\newblock J. Stat. Mech. 064004 (2016),
\newblock \doi{10.1088/1742-5468/2016/06/064004}.

\bibitem{Calabrese_RevQuenches}
P.~Calabrese and J.~Cardy,
\newblock \emph{Quantum quenches in 1+1 dimensional conformal field theories},
\newblock J. Stat. Mech. 064003 (2016),
\newblock \doi{10.1088/1742-5468/2016/06/064003}.

\bibitem{bernard2016conformal}
D.~Bernard and B.~Doyon,
\newblock \emph{Conformal field theory out of equilibrium: a review},
\newblock J. Stat. Mech. 064005 (2016) \doi{10.1088/1742-5468/2016/06/064005}.

\bibitem{Calabrese_IntroIntegrabilityDynamics}
P.~Calabrese, F.~H.~L. Essler and G.~Mussardo,
\newblock \emph{Introduction to `quantum integrability in out of equilibrium
  systems'},
\newblock J. Stat. Mech.  064001 (2016),
\newblock \doi{10.1088/1742-5468/2016/06/064001}.


\bibitem{caux-2013}
J.-S.~Caux and F.~H.~L.~Essler, {\it Time evolution of local observables after quenching to an integrable model},
{Phys. Rev. Lett. {\bf 110}, 257203 (2013)} \doi{10.1103/PhysRevLett.110.257203}.

\bibitem{caux-16}
J.-S. Caux, {\it The Quench Action},
{J. Stat. Mech. (2016) 064006}, \doi{10.1088/1742-5468/2016/06/064006}.

\bibitem{essler2016quench}
F.~H.~L. Essler and M.~Fagotti,
\newblock \emph{Quench dynamics and relaxation in isolated integrable quantum  spin chains},
\newblock J. Stat. Mech. 064002 (2016),
\newblock \doi{10.1088/1742-5468/2016/06/064002}.

\bibitem{sotiriadis-2012}
D. Fioretto and G. Mussardo, {\it Quantum quenches in integrable field theories},
{New J. Phys. {\bf 12}, 055015 (2010)} \doi{10.1088/1367-2630/12/5/055015};\\
S. Sotiriadis, D. Fioretto, and G. Mussardo, {\it Zamolodchikov-Faddeev algebra and quantum quenches in integrable field theories},
{J. Stat. Mech. (2012) P02017}, \doi{10.1088/1742-5468/2012/02/P02017}.


\bibitem{d-14}
G. Delfino, \emph{Quantum quenches with integrable pre-quench dynamics}, J. Phys. A {\bf 47} (2014) 402001, \doi{10.1088/1751-8113/47/40/402001};\\
G. Delfino and J. Viti, \emph{On the theory of quantum quenches in near-critical systems}, {J. Phys. A {\bf 50} (2017) 084004}, \doi{10.1088/1751-8121/aa5660}.

\bibitem{ra-20}
C. Rylands and N. Andrei, {\it Non-equilibrium aspects of integrable models},
Ann. Rev. Cond. Matt. Phys. {\bf 11}, 147 (2020) \doi{10.1146/annurev-conmatphys-031119-050630}.


\bibitem{alba2021reviewGHDinhomQuenches}
V.~Alba, B.~Bertini, M.~Fagotti, L.~Piroli and P.~Ruggiero,
\newblock \emph{Generalized-hydrodynamic approach to inhomohenous quenches:
  Correlations, entanglement and quantum effects},
\newblock to appear (2021).

\bibitem{Polkovnikov_ColloquiumNonEquilibrium}
A.~Polkovnikov, K.~Sengupta, A.~Silva and M.~Vengalattore,
\newblock \emph{Colloquium: Nonequilibrium dynamics of closed interacting
  quantum systems},
\newblock Rev. Mod. Phys. \textbf{83}, 863 (2011), \doi{10.1103/RevModPhys.83.863}.

\bibitem{Cazalilla_RevUltracold}
M.~A. Cazalilla, R.~Citro, T.~Giamarchi, E.~Orignac and M.~Rigol,
\newblock \emph{One dimensional bosons: From condensed matter systems to ultracold gases},
\newblock Rev. Mod. Phys. \textbf{83}, 1405 (2011), \doi{10.1103/RevModPhys.83.1405}.

\bibitem{Gogolin_ReviewIsolatedSystems}
C.~Gogolin and J.~Eisert,
\newblock \emph{Equilibration, thermalisation, and the emergence of statistical
  mechanics in closed quantum systems},
\newblock Rep. Prog. Phys. \textbf{79}, 056001 (2016),
\newblock \doi{10.1088/0034-4885/79/5/056001}.

\bibitem{eisert2015quantum}
J.~Eisert, M.~Friesdorf and C.~Gogolin,
\newblock \emph{Quantum many-body systems out of equilibrium},
\newblock Nature Phys. \textbf{11}, 124 (2015),
\newblock \doi{10.1038/nphys3215}.

\bibitem{dalessio2016quantum}
L.~D'Alessio, Y.~Kafri, A.~Polkovnikov and M.~Rigol,
\newblock \emph{From quantum chaos and eigenstate thermalization to statistical
  mechanics and thermodynamics},
\newblock Adv.  Phys. \textbf{65}, 239 (2016),
\newblock \doi{10.1080/00018732.2016.1198134}.

\bibitem{rv-21}
D. Rossini and E. Vicari, {\it Coherent and dissipative dynamics at quantum phase transitions},
arXiv:2103.02626 \eprint{2103.02626}.


\bibitem{greiner2002quantum}
M.~Greiner, O.~Mandel, T.~Esslinger, T.~W. H{\"a}nsch and I.~Bloch,
\newblock \emph{Quantum phase transition from a superfluid to a Mott insulator in a gas of ultracold atoms},
\newblock Nature \textbf{415}, 39 (2002) \doi{10.1038/415039a}.

\bibitem{Zwierlein_2005}
M.~W.~Zwierlein, C.~H.~Schunck, C.~A.~Stan, S.~M.~F.~Raupach and W.~Ketterle,
\newblock \emph{Formation Dynamics of a Fermion Pair Condensate},
\newblock Phys. Rev. Lett. \textbf{94}, 180401 (2005),
\newblock \doi{10.1103/PhysRevLett.94.180401}
  
  %

\bibitem{Hofferberth_2007}
S.~Hofferberth, I.~Lesanovsky, B.~Fischer, T.~ Schumm, and J.~Schmiedmayer,
\newblock \emph{Non-equilibrium coherence dynamics in one-dimensional Bose gases},
\newblock Nature \textbf{449}, 324 (2007),
\newblock \doi{10.1038/nature06149}.

\bibitem{kww-06} T. Kinoshita, T. Wenger, and D. S. Weiss, \emph{A quantum Newton cradle}, 
{Nature {\bf 440}, 900 (2006)}, \doi{10.1038/nature04693}.

\bibitem{Cheneau_2012}
M.~Cheneau, P.~Barmettler, D.~Poletti, M.~Endres, P.~Schau\ss, T.~Fukuhara, C.~Gross, I.~Bloch, C.~Kollath, Corinna and S.~Kuhr,
\newblock \emph{Light-cone-like spreading of correlations in a quantum many-body system},
\newblock Nature \textbf{481}, 484 (2012),
\newblock \doi{10.1038/nature10748}.


\bibitem{kaufman-2016}
A.~M.~Kaufman, M.~E.~Tai, A.~Lukin, M.~Rispoli, R.~Schittko, P.~M.~Preiss, and  M.~Greiner, 
{\it Quantum thermalisation through entanglement in an isolated many-body system}, 
Science {\bf 353}, 794  (2016), \doi{10.1126/science.aaf6725}.

\bibitem{brydges-2018}
T.~Brydges, A. Elben, P. Jurcevic, B.~Vermersch, C. Maier, B. P. Lanyon,  P. Zoller, R. Blatt, and C. F. Roos,
  \textit{Probing entanglement entropy via randomized measurements},
Science {\bf 364}, 260 (2019), \doi{10.1126/science.aau4963}.



\bibitem{Calabrese_QuenchesCorrelationsPRL}
P.~Calabrese and J.~Cardy,
\newblock \emph{Time dependence of correlation functions following a quantum quench},
\newblock Phys. Rev. Lett. \textbf{96}, 136801 (2006), \doi{10.1103/PhysRevLett.96.136801}.


\bibitem{Cazalilla_MasslessQuenchLL}
M.~A. Cazalilla,
\newblock \emph{Effect of suddenly turning on interactions in the Luttinger
  model},
\newblock Phys. Rev. Lett. \textbf{97}, 156403 (2006),
\newblock \doi{10.1103/PhysRevLett.97.156403}.

\bibitem{Calabrese_QuenchesCorrelationsLong}
P.~Calabrese and J.~Cardy,
\newblock \emph{Quantum quenches in extended systems}, J. Stat. Mech. P06008 (2007), \doi{10.1088/1742-5468/2007/06/P06008}.


\bibitem{Kollath_2007}
C.~Kollath, A.~M.~L\"auchli and E. Altman, 
\newblock \emph{Quench Dynamics and Nonequilibrium Phase Diagram of the Bose-Hubbard Model},
\newblock Phys. Rev. Lett. \textbf{98}, 180601 (2007),
\newblock \doi{10.1103/PhysRevLett.98.180601}

\bibitem{Rigol_2009}
M.~Rigol,
\newblock \emph{Quantum quenches and thermalization in one-dimensional fermionic systems},
\newblock Phys. Rev. A \textbf{80}, 053607 (2009)
\doi{10.1103/PhysRevA.80.053607}


\bibitem{piroli2017what}
L.~Piroli, B.~Pozsgay and E.~Vernier,
\newblock \emph{What is an integrable quench?},
\newblock Nucl. Phys. B \textbf{925}, 362 (2017),
\newblock \doi{10.1016/j.nuclphysb.2017.10.012}.

\bibitem{Haldane_LuttingerLiquid}
F.~D.~M. Haldane,
\newblock \emph{Effective harmonic-fluid approach to low-energy properties of
  one-dimensional quantum fluids},
\newblock Phys. Rev. Lett. \textbf{47}, 1840 (1981),
\newblock \doi{10.1103/PhysRevLett.47.1840}.

\bibitem{Luttinger_TLLiquid}
J.~M. Luttinger,
\newblock \emph{An exactly soluble model of a  many fermion system},
\newblock J. Math. Phys. \textbf{4}, 1154 (1963),
\newblock \doi{10.1063/1.1704046}.


\bibitem{iucci2009quantum}
A.~Iucci and M.~Cazalilla,
\newblock \emph{Quantum quench dynamics of the Luttinger model},
\newblock Phys. Rev. A \textbf{80}, 063619 (2009) \doi{10.1103/PhysRevA.80.063619}.

\bibitem{p-06}
E. Perfetto, {\it Time-dependent evolution of two coupled Luttinger liquids},
Phys. Rev. B {\bf 74},  205123 (2006) \doi{10.1103/PhysRevB.74.205123}.

\bibitem{ps-11}
E. Perfetto and G. Stefanucci, {\it On the thermalization of a Luttinger liquid after a sequence of sudden interaction quenches},
EPL {\bf 95}, 10006 (2011) \doi{10.1209/0295-5075/95/10006}.

\bibitem{iucci2010quantum}
A.~Iucci and M.~Cazalilla,
\newblock \emph{Quantum quench dynamics of the sine-Gordon model in some
  solvable limits},
\newblock New J. Phys. \textbf{12}, 055019 (2010), \doi{10.1088/1367-2630/12/5/055019}.

\bibitem{cazalilla2012thermalization}
M.~A. Cazalilla, A.~Iucci and M.-C. Chung,
\newblock \emph{Thermalization and quantum correlations in exactly solvable
  models},
\newblock Phys. Rev. E \textbf{85}, 011133 (2012), \doi{10.1103/PhysRevE.85.011133}.

\bibitem{dpf-13}
B. Dora, F. Pollmann, J. Fortagh, and G. Zarand, {\it Loschmidt echo and the many-body orthogonality catastrophe in a qubit-coupled Luttinger liquid},
Phys. Rev. Lett. {\bf 111}, 046402 (2013). \doi{10.1103/PhysRevLett.111.046402}.

\bibitem{rsm-12}
J. Rentrop, D. Schuricht, V. Meden, {\it Quench dynamics of the Tomonaga Luttinger model with momentum-dependent interaction},
New J. Phys. {\bf 14}, 075001 (2012) \doi{10.1088/1367-2630/14/7/075001}.

\bibitem{dbz-12}
B. Dora, A. Bacsi, and G. Zarand, {\it Generalized Gibbs ensemble and work statistics of a quenched Luttinger liquid}
Phys. Rev. B {\bf 86}, 161109 (2012) \doi{10.1103/PhysRevB.86.161109}.

\bibitem{bd-13}
A. Bacsi and B. Dora, {\it Quantum quench in the Luttinger model with finite temperature initial state},
Phys. Rev. B {\bf 88}, 155115 (2013), \doi{10.1103/PhysRevB.88.155115}.

\bibitem{ni-13}
N. Nessi and A. Iucci, {\it Quantum quench dynamics of the Coulomb Luttinger model}, Phys. Rev. B {\bf 87}, 085137 (2013), \doi{10.1103/PhysRevB.87.085137}.

\bibitem{nbm-13}
S. N. Dinh, D. A. Bagrets, and A. D. Mirlin, {\it Interaction quench in nonequilibrium Luttinger liquids},
Phys. Rev. B {\bf 88}, 245405 (2013), \doi{10.1103/PhysRevB.88.245405}.

\bibitem{kkm-14}
D.M. Kennes, C. Klockner, and V. Meden, {\it Spectral Properties of One-Dimensional Fermi Systems after an Interaction Quench},
Phys. Rev. Lett. {\bf 113}, 116401 (2014), \doi{10.1103/PhysRevLett.113.116401}.

\bibitem{bd-15}
M. Buchhold and S. Diehl, {\it Nonequilibrium universality in the heating dynamics of interacting Luttinger liquids},
Phys. Rev. A {\bf 92}, 013603 (2015), \doi{10.1103/PhysRevA.92.013603}.

\bibitem{dls-16}
B. Dora, R. Lundgren, M. Selover, and F. Pollmann,
{\it Momentum-Space Entanglement and Loschmidt Echo in Luttinger Liquids after a Quantum Quench},
Phys. Rev. Lett. {\bf 117}, 010603 (2016), \doi{10.1103/PhysRevLett.117.010603}.

\bibitem{cgc-18}
A. Calzona, F. Maria Gambetta, M. Carrega, F. Cavaliere, T. L. Schmidt, and M. Sassetti,
{\it Universal scaling of quench-induced correlations in a one-dimensional channel at finite temperature},
SciPost Phys. {\bf 4}, 023 (2018), \doi{10.21468/SciPostPhys.4.5.023}.

\bibitem{md-21}
C. P. Moca and B. Dora, {\it Universal conductance of a PT-symmetric Luttinger liquid after a quantum quench},
arXiv:2011.04561.


\bibitem{Bernier_2014}
J.-S.~Bernier, R.~ Citro, C.~Kollath and E.~Orignac,
\emph{Correlation Dynamics During a Slow Interaction Quench in a One-Dimensional Bose Gas},
\newblock Phys. Rev. Lett. \textbf{112}, 065301 (2014),
\newblock \doi{10.1103/PhysRevLett.112.065301}



\bibitem{uhrig2009interaction}
G.~S. Uhrig,
\newblock \emph{Interaction quenches of Fermi gases},
\newblock Phys. Rev. A \textbf{80}, 061602 (2009), \doi{10.1103/PhysRevA.80.061602}.

\bibitem{karrasch2012luttinger}
C.~Karrasch, J.~Rentrop, D.~Schuricht and V.~Meden,
\newblock \emph{Luttinger-liquid universality in the time evolution after an
  interaction quench},
\newblock Phys. Rev. Lett. \textbf{109}, 126406 (2012), \doi{10.1103/PhysRevLett.109.126406}.


\bibitem{coira2013quantum}
E.~Coira, F.~Becca and A.~Parola,
\newblock \emph{Quantum quenches in one-dimensional gapless systems},
\newblock Eur. Phys. J. B \textbf{86}, 55 (2013), \doi{10.1140/epjb/e2012-30978-y}.

\bibitem{hu-13}
S. A. Hamerla and G. S. Uhrig, {\it One-dimensional fermionic systems after interaction quenches and their description by bosonic field theories}
New J. Phys. {\bf 15}, 073012 (2013) \doi{10.1088/1367-2630/15/7/073012}.

\bibitem{phd-13}
F. Pollmann, M. Haque, and B. Dora, {\it Linear quantum quench in the Heisenberg XXZ chain: time dependent Luttinger model description of a lattice system},
Phys. Rev. B {\bf 87}, 041109 (2013), \doi{10.1103/PhysRevB.87.041109}.

\bibitem{km-13}
D. M. Kennes and V. Meden, {\it Luttinger liquid properties of the steady state after a quantum quench},
Phys. Rev. B {\bf 88}, 165131 (2013), \doi{10.1103/PhysRevB.88.165131}.

\bibitem{coser-2014}
A. Coser, E. Tonni, and P. Calabrese, {\it Entanglement negativity after a global quantum quench},
{J. Stat. Mech. P12017 (2014)}, \doi{10.1088/1742-5468/2014/12/P12017}.

\bibitem{svp-14}
S. Sorg, L. Vidmar, L. Pollet, and F. Heidrich-Meisner,
{\it Relaxation and thermalization in the one-dimensional Bose-Hubbard model: A case study for the interaction quantum quench from the atomic limit},
Phys. Rev. A {\bf 90}, 033606 (2014), \doi{10.1103/PhysRevA.90.033606}.

\bibitem{dp-15}
B. Dora and F. Pollmann, {\it Absence of orthogonality catastrophe after a spatially inhomogeneous interaction quench in Luttinger liquids},
Phys. Rev. Lett. {\bf 115}, 096403 (2015), \doi{10.1103/PhysRevLett.115.096403}.


\bibitem{collura2015quantum}
M.~Collura, P.~Calabrese and F.~H.~L. Essler,
\newblock \emph{Quantum quench within the gapless phase of the spin $1/2$
  heisenberg xxz spin chain},
\newblock Phys. Rev. B \textbf{92}, 125131 (2015),
\newblock \doi{10.1103/physrevb.92.125131}.

\bibitem{mitra2011mode}
A.~Mitra and T.~Giamarchi,
\newblock \emph{Mode-coupling-induced dissipative and thermal effects at long
  times after a quantum quench},
\newblock Phys. Rev. Lett. \textbf{107}, 150602 (2011),
\newblock \doi{10.1103/PhysRevLett.107.150602}.

\bibitem{mitra2012thermalization}
A.~Mitra and T.~Giamarchi,
\newblock \emph{Thermalization and dissipation in out-of-equilibrium quantum
  systems: A perturbative renormalization group approach},
\newblock Phys. Rev. B \textbf{85}, 075117 (2012),
\newblock \doi{0.1103/physrevb.85.075117}.

\bibitem{mitra2012time}
A.~Mitra,
\newblock \emph{Time evolution and dynamical phase transitions at a critical
  time in a system of one-dimensional bosons after a quantum quench},
\newblock Phys. Rev. Lett. \textbf{109}, 260601 (2012), \doi{10.1103/PhysRevLett.109.260601}.

\bibitem{mitra2013correlation}
A.~Mitra,
\newblock \emph{Correlation functions in the prethermalized regime after a
  quantum quench of a spin chain},
\newblock Phys. Rev. B \textbf{87}, 205109 (2013), \doi{10.1103/PhysRevB.87.205109}.

\bibitem{essler_Hubbard}
F.~H.~L. Essler, H.~Frahm, F.~Gohmann, A.~Klümper and V.~E. Korepin,
\newblock \emph{The One-Dimensional Hubbard Model},
\newblock Cambridge University Press,
\newblock \doi{10.1017/CBO9780511534843} (2005).

\bibitem{yang1967some}
C.~N. Yang,
\newblock \emph{Some exact results for the many-body problem in one dimension
  with repulsive delta-function interaction},
\newblock Phys. Rev. Lett. \textbf{19}, 1312 (1967),
\newblock \doi{10.1103/PhysRevLett.19.1312}.

\bibitem{gaudin1967systeme}
M.~Gaudin,
\newblock \emph{Un systeme a une dimension de fermions en interaction},
\newblock Phys. Lett. A \textbf{24}, 55 (1967), \doi{10.1016/0375-9601(67)90193-4}.

\bibitem{Andrews_FirstExperimentInterferenceCondensates}
M.~R. Andrews, C.~G. Townsend, H.-J. Miesner, D.~S. Durfee, D.~M. Kurn and
  W.~Ketterle,
\newblock \emph{Observation of interference between two Bose condensates},
\newblock Science \textbf{275}, 637 (1997),
\newblock \doi{10.1126/science.275.5300.637}.

\bibitem{Shin_ExperimentSplitCondensates1}
Y.~Shin, M.~Saba, T.~A. Pasquini, W.~Ketterle, D.~E. Pritchard and A.~E.
  Leanhardt,
\newblock \emph{Atom interferometry with Bose-Einstein condensates in a
  double-well potential},
\newblock Phys. Rev. Lett. \textbf{92}, 050405 (2004),
\newblock \doi{10.1103/PhysRevLett.92.050405}.

\bibitem{Shin_ExperimentSplitCondensates2}
Y.~Shin, C.~Sanner, G.-B. Jo, T.~A. Pasquini, M.~Saba, W.~Ketterle, D.~E.
  Pritchard, M.~Vengalattore and M.~Prentiss,
\newblock \emph{Interference of Bose-Einstein condensates split with an atom
  chip},
\newblock Phys. Rev. A \textbf{72}, 021604 (2005),
\newblock \doi{10.1103/PhysRevA.72.021604}.

\bibitem{Shin_ExperimentSplitCondensates3}
G.-B. Jo, Y.~Shin, S.~Will, T.~A. Pasquini, M.~Saba, W.~Ketterle, D.~E.
  Pritchard, M.~Vengalattore and M.~Prentiss,
\newblock \emph{Long phase coherence time and number squeezing of two
  Bose-Einstein condensates on an atom chip},
\newblock Phys. Rev. Lett. \textbf{98}, 030407 (2007),
\newblock \doi{10.1103/PhysRevLett.98.030407}.

\bibitem{Schumm_ExperimentMatterWaveInterferometry}
T.~Schumm, S.~Hofferberth, L.~M. Andersson, S.~Wildermuth, S.~Groth,
  I.~Bar-Joseph, J.~Schmiedmayer and P.~Kr{\"u}ger,
\newblock \emph{Matter-wave interferometry in a double well on an atom chip},
\newblock Nature Phys. \textbf{1}, 57 (2005), \doi{10.1038/nphys125}.

\bibitem{Albiez_ObservationTunnellingBosonicJosephsonJunction}
M.~Albiez, R.~Gati, J.~F\"olling, S.~Hunsmann, M.~Cristiani and M.~K.
  Oberthaler,
\newblock \emph{Direct observation of tunneling and nonlinear self-trapping in
  a single bosonic Josephson junction},
\newblock Phys. Rev. Lett. \textbf{95}, 010402 (2005),
\newblock \doi{10.1103/PhysRevLett.95.010402}.

\bibitem{Gati_RealizationBosonicJosephsonJunction}
R.~Gati, M.~Albiez, J.~F{\"o}lling, B.~Hemmerling and M.~K. Oberthaler,
\newblock \emph{Realization of a single Josephson junction for Bose--Einstein
  condensates},
\newblock Appl. Phys. B \textbf{82}, 207 (2006), \doi{10.1007/s00340-005-2059-z}.

\bibitem{Levy_ExperimentalBosonicJosephsonEffects}
S.~Levy, E.~Lahoud, I.~Shomroni and J.~Steinhauer,
\newblock \emph{The A.C. and D.C. Josephson effects in a Bose--Einstein condensate},
\newblock Nature \textbf{449}, 579 (2007), \doi{10.1038/nature06186}.

\bibitem{Kuhnert_ExpEmergenceCharacteristicLength1d}
M.~Kuhnert, R.~Geiger, T.~Langen, M.~Gring, B.~Rauer, T.~Kitagawa, E.~Demler,
  D.~Adu~Smith and J.~Schmiedmayer,
\newblock \emph{Multimode dynamics and emergence of a characteristic length
  scale in a one-dimensional quantum system},
\newblock Phys. Rev. Lett. \textbf{110}, 090405 (2013),
\newblock \doi{10.1103/PhysRevLett.110.090405}.

\bibitem{kleine2008spin}
A.~Kleine, C.~Kollath, I.~P. McCulloch, T.~Giamarchi and U.~Schollw\"ock,
\newblock \emph{Spin-charge separation in two-component Bose gases},
\newblock Phys. Rev. A \textbf{77}, 013607 (2008),
\newblock \doi{10.1103/PhysRevA.77.013607}.

\bibitem{jk-15}
A. J. A. James and R. M. Konik, {\it Quantum quenches in two spatial dimensions using chain array matrix product states},
Phys. Rev. B {\bf 92}, 161111 (2015), \doi{10.1103/PhysRevB.92.161111}.

\bibitem{mk-08}
M. Moeckel and S. Kehrein, {\it Interaction Quench in the Hubbard model},
{Phys. Rev. Lett. {\bf 100}, 175702 (2008)}, \doi{10.1103/PhysRevLett.100.175702}.

\bibitem{ekw-09}
M. Eckstein, M. Kollar, and P. Werner,
{\it Interaction quench in the Hubbard model: Relaxation of the spectral function and the optical conductivity},
{Phys. Rev. B {\bf 81}, 115131 (2010)}, \doi{10.1103/PhysRevB.81.115131}.

\bibitem{qkns-14}
F. Queisser, K. V. Krutitsky, P. Navez, and R. Schutzhold, {\it Equilibration and prethermalization in the Bose-Hubbard and Fermi-Hubbard models},
{Phys. Rev. A {\bf 89}, 033616 (2014)}, \doi{10.1103/PhysRevA.89.033616}.

\bibitem{ymwr-14}
D. Iyer, R. Mondaini, S. Will, and M. Rigol,
{\it Coherent quench dynamics in the one-dimensional Fermi-Hubbard model},
{Phys. Rev. A {\bf 90}, 031602(R) (2014)}, \doi{10.1103/PhysRevA.90.031602}.

\bibitem{roh-15}
L. Riegger, G. Orso, and F. Heidrich-Meisner,
{\it Interaction quantum quenches in the one-dimensional Fermi-Hubbard model with spin imbalance},
{Phys. Rev. A {\bf 91}, 043623 (2015)}, \doi{10.1103/PhysRevA.91.043623}.


\bibitem{yr-17}
X. Yin and L. Radzihovsky,
{\it Quench dynamics of spin-imbalanced Fermi-Hubbard model in one dimension},
Phys. Rev. A {\bf 94}, 063637 (2016), \doi{10.1103/PhysRevA.94.063637}.

\bibitem{sjhb-17}
N. Schluenzen, J.-P. Joost, F. Heidrich-Meisner, and M. Bonitz,
{\it Nonequilibrium dynamics in the one-dimensional Fermi-Hubbard model: A comparison of the nonequilibrium Green functions approach and the density matrix renormalization group method},
{Phys. Rev. B {\bf 95}, 165139 (2017)}, \doi{10.1103/PhysRevB.95.165139}.

\bibitem{zvr-19}
Y. Zhang, L. Vidmar, and M. Rigol, {\it Quantum dynamics of impenetrable SU(N) fermions in one-dimensional lattices},
Phys. Rev. A {\bf 99}, 063605 (2019), \doi{10.1103/PhysRevA.99.063605}.

\bibitem{amlz-20}
M. Antal Werner, C. P. Moca, O. Legeza, and G. Zarand,
{\it Quantum Quench and Charge Oscillations in the SU(3) Hubbard Model: a Test of Time Evolving Block Decimation with general non-Abelian Symmetries},
Phys. Rev. B {\bf 102}, 155108 (2020), \doi{10.1103/PhysRevB.102.155108}.


\bibitem{piroli2019integrableI}
L.~Piroli, E.~Vernier, P.~Calabrese and B.~Pozsgay,
\newblock \emph{Integrable quenches in nested spin chains I: the exact steady
  states},
\newblock J. Stat. Mech. 063103 (2019),
\newblock \doi{10.1088/1742-5468/ab1c51}.

\bibitem{piroli2019integrableII}
L.~Piroli, E.~Vernier, P.~Calabrese and B.~Pozsgay,
\newblock \emph{Integrable quenches in nested spin chains II: fusion of
  boundary transfer matrices},
\newblock J. Stat. Mech. 063104 (2019),
\newblock \doi{10.1088/1742-5468/ab1c52}.

\bibitem{mestyan2017exact}
M.~Mestyán, B.~Bertini, L.~Piroli and P.~Calabrese,
\newblock \emph{Exact solution for the quench dynamics of a nested integrable system},
\newblock J. Stat. Mech. 083103 (2017),
\newblock \doi{10.1088/1742-5468/aa7df0}.


\bibitem{btc-17}
B. Bertini, E. Tartaglia, and P. Calabrese,
{\it Quantum Quench in the Infinitely Repulsive Hubbard Model: The Stationary State},
J. Stat. Mech. (2017) 103107, \doi{10.1088/1742-5468/aa8c2c}.

\bibitem{id-17}
E. Ilievski and J. De Nardis, {\it Ballistic transport in the one-dimensional Hubbard model: the hydrodynamic approach},
Phys. Rev. B {\bf 96}, 081118 (2017) \doi{10.1103/PhysRevB.96.081118}.

\bibitem{sf-10}
M. Schir\'o and M. Fabrizio, {\it Time-Dependent Mean Field Theory for Quench Dynamics in correlated electron systems},
{Phys. Rev. Lett. {\bf 105}, 076401 (2010)}, \doi{10.1103/PhysRevLett.105.076401};\\
M. Schir\'o and M. Fabrizio,
{\it Quantum Quenches in the Hubbard Model: Time Dependent Mean Field Theory and The Role of Quantum Fluctuations},
{Phys. Rev. B {\bf 83}, 165105 (2011)}, \doi{10.1103/PhysRevB.83.165105}.



\bibitem{robinson2016motion}
N.~J. Robinson, J.-S. Caux and R.~M. Konik,
\newblock \emph{Motion of a distinguishable impurity in the Bose gas: Arrested
  expansion without a lattice and impurity snaking},
\newblock Phys. Rev. Lett. \textbf{116}, 145302 (2016),
\newblock \doi{10.1103/PhysRevLett.116.145302}.

\bibitem{robinson2019light}
N.~J. Robinson, J.-S. Caux and R.~M. Konik,
\emph{Light cone dynamics in excitonic states of two-component Bose and Fermi gases}
J. Stat. Mech.  013103 (2020), \doi{10.1088/1742-5468/ab5706}


\bibitem{mestyan2019spin}
M.~Mestyán, B.~Bertini, L.~Piroli and P.~Calabrese,
\newblock \emph{Spin-charge separation effects in the low-temperature transport
  of one-dimensional fermi gases},
\newblock Phys. Rev. B \textbf{99}, 014305 (2019),
\newblock \doi{10.1103/physrevb.99.014305}.

\bibitem{vanNieuwkerk_QuenchSineGordonSelfConsistentHarmonicApprox}
Y.~D. van Nieuwkerk and F.~H.~L. Essler,
\newblock \emph{Self-consistent time-dependent harmonic approximation for the
  sine-gordon model out of equilibrium},
\newblock J. Stat. Mech. 084012 (2019),
\newblock \doi{10.1088/1742-5468/ab3579}.

\bibitem{vanNieuwkerk_TunnelCoupledBoseGasesLowEnergy}
Y.~D. van Nieuwkerk and F.~H.~L. Essler,
\newblock \emph{On the low-energy description for tunnel-coupled
  one-dimensional bose gases} (2020), \eprint{2003.07873}.

\bibitem{van2018projective}
Y.~D. Van~Nieuwkerk, J.~Schmiedmayer and F.~Essler,
\newblock \emph{Projective phase measurements in one-dimensional Bose gases},
\newblock SciPost Phys. \textbf{5}, 046 (2018), \doi{10.21468/SciPostPhys.5.5.046}.

\bibitem{vannieuwkerk2020josephson}
Y.~D. van Nieuwkerk, J.~Schmiedmayer and F.~H.~L. Essler,
\newblock \emph{Josephson oscillations in split one-dimensional Bose gases}
  (2020), \eprint{2010.11214}.

\bibitem{Gring_Prethermalization}
M.~Gring, M.~Kuhnert, T.~Langen, T.~Kitagawa, B.~Rauer, M.~Schreitl, I.~Mazets,
  D.~A. Smith, E.~Demler and J.~Schmiedmayer,
\newblock \emph{Relaxation and prethermalization in an isolated quantum
  system},
\newblock Science \textbf{337}, 1318 (2012), \doi{10.1126/science.1224953}.

\bibitem{Smith_PrethermalizationFromFullDistributions}
D.~A. Smith, M.~Gring, T.~Langen, M.~Kuhnert, B.~Rauer, R.~Geiger, T.~Kitagawa,
  I.~Mazets, E.~Demler and J.~Schmiedmayer,
\newblock \emph{Prethermalization revealed by the relaxation dynamics of full
  distribution functions},
\newblock New J. Phys. \textbf{15}, 075011 (2013), \doi{10.1088/1367-2630/15/7/075011}.

\bibitem{Foini_CoupledLLSchmiedtmayerMassive}
L.~Foini and T.~Giamarchi,
\newblock \emph{Relaxation dynamics of two coherently coupled one-dimensional
  bosonic gases},
\newblock Eur. Phys. J. Spec. Top. \textbf{226}, 2763 (2017), \doi{10.1140/epjst/e2016-60383-x}.

\bibitem{Foini_CoupledLLsMassiveMassless}
L.~Foini and T.~Giamarchi,
\newblock \emph{Nonequilibrium dynamics of coupled luttinger liquids},
\newblock Phys. Rev. A \textbf{91}, 023627 (2015), \doi{10.1103/PhysRevA.91.023627}.

\bibitem{kormos2016quantum}
M.~Kormos and G.~Zar{\'a}nd,
\newblock \emph{Quantum quenches in the sine-gordon model: a semiclassical approach},
\newblock Phys. Rev. E \textbf{93}, 062101 (2016), doi{10.1103/PhysRevE.93.062101}.

\bibitem{horvath2019nonequilibrium}
D.~Horv{\'a}th, I.~Lovas, M.~Kormos, G.~Tak{\'a}cs and G.~Zar{\'a}nd,
\newblock \emph{Nonequilibrium time evolution and rephasing in the quantum
  sine-gordon model},
\newblock Phys. Rev. A \textbf{100}, 013613 (2019), \doi{10.1103/PhysRevA.100.013613}.

\bibitem{Langen_UnequalLL}
T.~Langen, T.~Schweigler, E.~Demler and J.~Schmiedmayer,
\newblock \emph{Double light-cone dynamics establish thermal states in
  integrable 1d bose gases},
\newblock New J. Phys. \textbf{20}, 023034 (2018), \doi{10.1088/1367-2630/aaaaa5}.

\bibitem{Kitagawa_DynamicsPrethermalizationQuantumNoise}
T.~Kitagawa, A.~Imambekov, J.~Schmiedmayer and E.~Demler,
\newblock \emph{The dynamics and prethermalization of one-dimensional quantum
  systems probed through the full distributions of quantum noise},
\newblock New J. Phys. \textbf{13}, 073018 (2011), \doi{10.1088/1367-2630/13/7/073018}.

\bibitem{ruggiero2021large}
P.~Ruggiero, L.~Foini and T.~Giamarchi,
\newblock \emph{Large-scale thermalization, prethermalization, and impact of
  temperature in the quench dynamics of two unequal Luttinger liquids},
\newblock Phys. Rev. Res. \textbf{3}, 013048 (2021),
\newblock \doi{10.1103/physrevresearch.3.013048}.

\bibitem{kardar1986josephson}
M.~Kardar
\newblock \emph{Josephson-junction ladders and quantum fluctuations},
\newblock Phys. Rev. B \textbf{33}, 3125 (1986),
\newblock \doi{10.1103/PhysRevB.33.3125}.

\bibitem{gritsev2007linear}
V.~Gritsev, A.~Polkovnikov, E.~Demler
\newblock \emph{Linear response theory for a pair of coupled one-dimensional condensates of interacting atoms},
\newblock Phys. Rev. B \textbf{75}, 174511 (2007),
\newblock \doi{10.1103/PhysRevB.75.174511}.

\bibitem{bkc-14}
L. Bucciantini, M. Kormos, and P. Calabrese,
{\it Quantum quenches from excited states in the Ising chain},
{J. Phys. A {\bf 47}, 175002 (2014)} \doi{10.1088/1751-8113/47/17/175002}.

\bibitem{diehl-86}
H. W. Diehl, 
{\it The theory of boundary critical phenomena Phase Transitions and Critical Phenomena},
{vol 10, Edition C. Domb and J.L. Lebowitz (1986)}.

\bibitem{sotiriadis2014validity}
S.~Sotiriadis, P.~Calabrese,
 \emph{Validity of the GGE for quantum quenches from interacting to noninteracting models},
{J. Stat. Mech. (2014) P07024}, \doi{10.1088/1742-5468/2014/07/P07024}

\bibitem{horvath2016initial}
D.X.~Horvath, S. Sotiriadis, G. Tak\'acs,
\newblock \emph{Initial states in integrable quantum field theory quenches from an integral equation hierarchy}
\newblock Nucl. Phys. B \textbf{902} (2016), \doi{10.1016/j.nuclphysb.2015.11.025}

\bibitem{kst-18}
I. Kukuljan, S. Sotiriadis, and G. Takacs, {\it Correlation Functions of the Quantum Sine-Gordon Model in and out of Equilibrium},
 Phys. Rev. Lett. {\bf 121}, 110402 (2018), \doi{10.1103/PhysRevLett.121.110402}.

\bibitem{kukuljan2020out}
I.~Kukuljan, S.~Sotiriadis, G.~Tak\'acs, 
\newblock \emph{Out-of-horizon correlations following a quench in a relativistic quantum field theory},
\newblock JHEP {\bf 7} (2020),  \doi{10.1007/JHEP07(2020)224}

\bibitem{gaussification}
T.~Schweigler, M.~Gluza, M.~Tajik, S.~Sotiriadis, F.~Cataldini, S.-C.~Ji, F.S.~Møller, J.~Sabino, B.~Rauer, J.~Eisert, J.~Schmiedmayer
\newblock \emph{Decay and recurrence of non-Gaussian correlations in a quantum many-body system},
\newblock Nature Physiscs \textbf{1-5} (2021), \doi{10.1038/s41567-020-01139-2}.


\bibitem{cardy2016further}
J.~Cardy,
\newblock \emph{Quantum quenches to a critical point in one dimension: some
  further results},
\newblock J. Stat. Mech. 023103 (2016),
\newblock \doi{10.1088/1742-5468/2016/02/023103}.

\bibitem{sc-08}
S. Sotiriadis and J. Cardy, {\it Inhomogeneous Quantum Quenches},
{J. Stat. Mech. (2008) P11003}, \doi{10.1088/1742-5468/2008/11/P11003}.

\bibitem{c-14}
J. Cardy, {\it Thermalization and Revivals after a Quantum Quench in Conformal Field Theory},
Phys. Rev. Lett. {\bf 112}, 220401 (2014), \doi{10.1103/PhysRevLett.112.220401}.

\bibitem{sc-10}
S. Sotiriadis and J. Cardy, {\it Quantum quench in interacting field theory: a self-consistent approximation},
Phys. Rev. B {\bf 81}, 134305 (2010), \doi{10.1103/PhysRevB.81.134305}.

\bibitem{gc-11}
A. Gambassi and P. Calabrese, {\it Quantum quenches as classical critical films}, EPL {\bf 95}, 66007 (2011), \doi{10.1209/0295-5075/95/66007}.

\bibitem{dsvc-16}
J. Dubail, J.-M. St\'ephan, J. Viti, and P. Calabrese,
{\it Conformal Field Theory for Inhomogeneous One-dimensional Quantum Systems: the Example of Non-Interacting Fermi Gases},
{SciPost Phys. {\bf 2}, 002 (2017)}, \doi{10.21468/SciPostPhys.2.1.002}.

\bibitem{diehl1997theory}
H.~W. Diehl,
\newblock \emph{The theory of boundary critical phenomena},
\newblock Int. J. Mod. Phys. B \textbf{11}, 3503 (1997) \doi{10.1142/S0217979297001751}.

\bibitem{stm-14}
S. Sotiriadis, G. Takacs, and G. Mussardo, {\it Boundary State in an Integrable Quantum Field Theory Out of Equilibrium},
Phys. Lett. B {\bf 734}, 52 (2014), \doi{10.1016/j.physletb.2014.04.058}.

\bibitem{hst-16}
D. X. Horvath, S. Sotiriadis, and G. Takacs, {\it Initial states in integrable quantum field theory quenches from an integral equation hierarchy},
Nucl. Phys. B {\bf 902}, 508 (2016), \doi{10.1016/j.nuclphysb.2015.11.025}.

\bibitem{cc-05}
P. Calabrese and J. Cardy, {\it Evolution of entanglement entropy in one-dimensional systems},
{J. Stat. Mech. (2005) P04010}, \doi{10.1088/1742-5468/2005/04/P04010}.

\bibitem{carleo2014lightcone}
G.~Carleo, F.~Becca, L.~Sanchez-Palencia, S.~Sorella and M.~Fabrizio,
\newblock \emph{Light-cone effect and supersonic correlations in one- and
  two-dimensional bosonic superfluids},
\newblock Phys. Rev. A \textbf{89}, 031602 (2014),
\newblock \doi{10.1103/PhysRevA.89.031602}.

\bibitem{bonnes2014lightcone}
L.~Bonnes, F.~H.~L. Essler and A.~M. L\"auchli,
\newblock \emph{``Light-cone'' dynamics after quantum quenches in spin chains},
\newblock Phys. Rev. Lett. \textbf{113}, 187203 (2014),
\newblock \doi{10.1103/PhysRevLett.113.187203}.

\bibitem{cbp-12}
M. Cheneau, P. Barmettler, D. Poletti, M. Endres, P. Schau{\ss}, T. Fukuhara, C. Gross, I. Bloch, C. Kollath, and S. Kuhr,
{\it Light-cone-like spreading of correlations in a quantum many-body system},
{Nature {\bf 481}, 484 (2012)}, \doi{10.1038/nature10748}


\bibitem{geiger2014local}
R.~Geiger, T.~Langen, I.~Mazets and J.~Schmiedmayer,
\newblock \emph{Local relaxation and light-cone-like propagation of
  correlations in a trapped one-dimensional Bose gas},
\newblock New J. Phys. \textbf{16}, 053034 (2014), \doi{10.1088/1367-2630/16/5/053034}.

\bibitem{dsc-17}
J. Dubail, J.-M. Stephan, and P. Calabrese, {\it Emergence of curved light-cones in a class of inhomogeneous Luttinger liquids},
{SciPost Phys. {\bf 3}, 019 (2017)} \doi{10.21468/SciPostPhys.3.3.019}.

\bibitem{Rigol_GGE}
M.~Rigol, V.~Dunjko, and M.~Olshanii,
\newblock \emph{Thermalization and its mechanism for generic isolated quantum systems},
\newblock Nature \textbf{452}, 854 (2008), \doi{10.1038/nature06838}.


\bibitem{Bachas_PermeableWalls}
C.~Bachas, J.~de~Boer, R.~Dijkgraaf and H.~Ooguri,
\newblock \emph{Permeable conformal walls and holography},
\newblock JHEP \textbf{06}, 027 (2002), \doi{10.1088/1126-6708/2002/06/027}.


\bibitem{unfolded_coupledCFT}
P.~Ruggiero, P.~Calabrese, T.~Giamarchi and L.~Foini,
\newblock \emph{In preparation.}

\bibitem{oshikawa1996defect}
M.~Oshikawa, I.~Affleck,
\newblock \emph{Defect Lines in the Ising Model and Boundary States on Orbifolds},
\newblock Phys. Rev. Lett. \textbf{77}, 2604 (1996).
\newblock \doi{10.1103/PhysRevLett.77.2604}.

\bibitem{oshikawa1997boundary}
M.~Oshikawa, I.~Affleck,
\newblock \emph{Boundary conformal field theory approach to the critical two-dimensional Ising model with a defect line},
\newblock Nucl. Phys. B \textbf{495} 533 (1997).
\newblock \doi{10.1016/S0550-3213(97)00219-8}.



\bibitem{bachas2008fusion}
C.~Bachas and I.~Brunner,
\newblock \emph{Fusion of conformal interfaces},
\newblock JHEP \textbf{06}, 085 (2008).

\bibitem{bms-07}
B. Bellazzini, M. Mintchev, and P. Sorba, {\it Bosonization and Scale Invariance on Quantum Wires},
J. Phys. A {\bf 40}, 2485 (2007) \doi{10.1088/1751-8113/40/10/017}, \doi{10.1088/1126-6708/2008/02/085}.


\bibitem{yellowbook}
P.~Di~Francesco, P.~Mathieu and D.~Senechal,
\newblock \emph{Conformal Field Theory},
\newblock Springer, New York, USA (1997), \doi{10.1007/978-1-4612-2256-9}.

%\bibitem{Wen_Entanglement_interface}
%X.~Wen, Y.~Wang and S.~Ryu,
%\newblock \emph{Entanglement evolution across a conformal interface},
%\newblock J. Phys. A \textbf{51}, 195004 (2018).


\end{thebibliography}
\end{document}